\documentclass[aoas,preprint]{imsart}

\RequirePackage{amsthm,amsmath,amsfonts,amssymb}
\RequirePackage[authoryear]{natbib}
\RequirePackage[colorlinks,citecolor=blue,urlcolor=blue]{hyperref}
\RequirePackage{graphicx}

\startlocaldefs
\theoremstyle{plain}

\theoremstyle{remark}

\endlocaldefs

\usepackage{xfp}
\newcommand\SupplementaryMaterials{%
  \xdef\presupfigures{\arabic{figure}}
  \xdef\presuptables{\arabic{table}}
  \xdef\presupsections{\arabic{section}}
  \renewcommand\thefigure{A\fpeval{\arabic{figure}-\presupfigures}}
  \renewcommand\thetable{A\fpeval{\arabic{table}-\presuptables}}
}

\begin{document}
\begin{changemargin}{-1cm}{-9mm}{-5mm}
\setlength{\textheight}{8.6in}
\setlength{\paperheight}{279mm}
\setlength{\paperwidth}{216mm}

\begin{frontmatter}
\title{Model-based calibration of gear-specific fish abundance survey data as a change-of-support problem}
\runtitle{Gear calibration as COSP}

\begin{aug}
\author[A,B,C]{\fnms{Grace S.} \snm{Chiu}\ead[label=e1]{gschiu@vims.edu}},
\author[D,C]{\fnms{Anton H.} \snm{Westveld}},
\author[E]{\fnms{Mark A.} \snm{Albins}},
\author[F]{\fnms{Kevin M.} \snm{Boswell}},
\author[A]{\fnms{John M.} \snm{Hoenig}},
\author[E]{\fnms{Sean P.} \snm{Powers}},
\author[G]{\fnms{S.~Lynne} \snm{Stokes}}
\and
\author[H]{\fnms{Allison L.} \snm{White}}
\address[A]{Batten School of Coastal \& Marine Sciences, William \& Mary  / Virginia Institute of Marine Science}
\address[B]{Department of Statistical Sciences and Operations Research, Virginia Commonwealth University}
\address[C]{Department of Statistics, University of Washington; Department of Statistics and Actuarial Science, University of Waterloo}
\address[D]{Research School of Finance, Actuarial Studies and Statistics, Australian National University}
\address[E]{Stokes School of Marine and Environmental Sciences, University of South Alabama}
\address[F]{Institute of Environment \& Department of Biological Sciences, Florida International University}
\address[G]{Department of Statistical Science, Southern Methodist University}
\address[H]{Southeast Fisheries Science Center, NOAA}
\address[]{Corresponding author: \printead{e1}}
\end{aug}

\begin{abstract}
For commercial and recreational fisheries of a wide-ranging species to be sustainable, abundance studies from neighboring regions should be unified. For the first time in the USA, a single research project to estimate the abundance of the Greater Amberjack (\textit{Seriola dumerili}) is being undertaken at the continental scale. A major methodological challenge lies in 1) the difference in fish detection gears deployed by regional survey teams that produce gear-specific relative abundance indices, and 2) the unknown relationship between actual abundance and these indices. In this paper, we develop a conversion tool that is operationalized from a Bayesian hierarchical model in an inferential context akin to the change-of-support problem often encountered in large-scale spatial studies; though, the context here is to reconcile abundance data observed at various gear-specific scales. To this end, we consider a small calibration experiment in which 2 to 4 different underwater video camera types were simultaneously deployed on each of 21 boat trips. Alongside the suite of deployed cameras was also an acoustic echosounder that recorded fish signals along surrounding transects. Our modeling framework is used to derive calibration formulae for translating camera-specific relative indices to the actual abundance scale in surveys that deploy a single camera. Cross-validation is conducted using mark-recapture abundance estimates (only available for 10 trips, all observed at a single habitat type) and through a separate simulation study. We also briefly discuss the case when surveys pair one camera with the echosounder. 

\end{abstract}


\end{frontmatter}

\section{INTRODUCTION}\label{sec-intro}

For the first time in the USA, a single research project to estimate the abundance of the fish species \emph{Seriola dumerili} (common name: Greater Amberjack, or GAJ) is being undertaken at the continental scale, stretching across the South Atlantic and Gulf of Mexico regions. This is referred to as the Greater Amberjack Research Study. Prior to this effort, smaller studies have shed light on the abundance of various species in regional waters based on relative abundance indices derived from surveys that typically employ underwater video cameras. However, camera types may differ across regions, e.g., murkier regions require a camera type that is better suited for low visibility. Moreover, camera-based relative abundance data are typically in the form of a MaxN count, whereby video footage is broken into frames, each frame yielding a count of the species of interest, and MaxN is taken to be the maximum of this count across all frames. With its unique field of view and associated video processing protocol, each camera type therefore produces gear-specific relative abundance counts that (1) does not reflect absolute abundance (an actual count of individuals present), and (2) are only meaningful when compared within gear type but not between. 

\cite{Campbell2015} investigated the relationship between MaxN
and mean absolute abundance for a given camera gear, and \cite{Campbell2018}, that between MaxN's from various camera gears. However, the purpose of
these works was to explore ways in operational settings that can make
better use of camera gears to reflect absolute abundance. Instead, our goal is to translate camera-specific MaxN counts to the scale of absolute abundance. 

Through a small-scale survey, \cite{Scoulding2023} demonstrated that paired deployment of a remote underwater video (RUV) camera alongside an acoustic echosounder allows each gear type to compensate for the other's limitations, leading to a more reliable abundance estimate than one based on a single gear type. In particular, while the echosounder records acoustic signals of fish on the absolute scale, it is highly subject to confounding from acoustically similar species, thus requiring an externally derived correction factor. Conversely, a camera allows visual distinction between species but records data on a relative scale that is specific to the camera type. 

The paired design is employed in most regional surveys of the continental-scale GAJ Research Study. However, some of the surveys deploy a single camera type, and others, the acoustic echosounder alone, depending on the region and habitat being surveyed. To consolidate all regional surveys for the estimation of a study-wide absolute abundance, all regional datasets must be reconciled to represent regional absolute abundances. Thus, we are faced with the following operational challenges:

\vspace{5mm}

\begin{enumerate}
\def\labelenumi{\alph{enumi}.}
\item
  How can we reconcile the inherent difference in representation between a camera-based MaxN (camera-specific relative abundance of GAJ) and an acoustic count (absolute abundance of GAJ and GAJ-like species), given that each of the two counts is inherently
  related to the same true abundance in its unique way?
\item
  How do we ensure that the conversion from MaxN to absolute abundance for any specific camera type is done in a unified manner that allows us to derive a meaningful study-wide total abundance based on regional conversions that involve different camera types?
\item
  How can we estimate absolute abundance with acoustic data only, without camera data to correct for the confounding of acoustically indistinguishable species?\\
\end{enumerate}

In this paper, we simultaneously address the challenges (a) and (b) whenever a camera is deployed, either alone or paired with an acoustic echosounder. (Challenge (c) will be handled outside of this paper.) This requires an understanding of the joint relationship among all gear-specific MaxN counts and the acoustic count. To do so, we conduct a small-scale calibration experiment in the Alabama-Mississippi region (one of four regions that comprise the full-scale GAJ Research Study), in which 21 separate boat trips surveyed 13 unique reef structures, each trip deploying the acoustic echosounder alongside two to all of four types of cameras: drop camera, stereo baited RUV (SBRUV), trap camera, and remotely operated vehicle (ROV) camera. For these experimental data, we develop a cross-calibration
modeling framework through a
single Bayesian hierarchical model that jointly relates all available MaxN data to acoustic data. The unifying
calibration methodology produces a joint regression fit with confidence
bands for each camera-specific MaxN. The calibrated value and associated uncertainty estimate produced as such
simultaneously accounts for all camera types and the acoustic gear under
a single overarching model. 

Our methodological approach to unify multiple sources of
data to make inference for an underlying ``truth'' is similar to those
in \cite{Chiu2011a, Chiu2013, Chiu2013b}, and \cite{Kuh2023a}, all of which exploit the
hierarchical model structure to address a challenge that is akin to the
change-of-support problem (COSP). Although COSP is typically viewed as
reconciling spatial data observed at multiple spatial scales \citep{Gelfand2010, GotwayCrawford}, in this paper, we consider
a wide-sense COSP in the context of reconciling abundance data observed
at various gear-specific scales, some being relative, and others,
absolute. 

Note that unlike actual surveys in the GAJ Research Study which deployed at most one camera type
(with or without acoustic gear), the 21 data points from the
calibration experiment (with data from the echosounder and at least two of four camera types) play a crucial role in addressing challenges (a)
and (b) stated above. Moreover, the model's hierarchical nature allows missing data
to be internally imputed, so that on boat trips that have fewer than four camera types, the Bayesian paradigm allows imputation of missing observations through the joint inference of all unobserved quantities in the model.  

The rest of this paper is structured as follows. Section~\ref{sec-calibdata} describes some key features of the data from the calibration experiment. Section~\ref{sec-framework} delves into our calibration modeling framework, associated rationale, assumptions, caveats, and results. Supplementary details for both sections appear in the Appendix. In Section~\ref{sec-formulae}, we develop a procedure that operationalizes the framework from Section~\ref{sec-framework} in the case of a camera that is deployed alone (unpaired with the echosounder) in the field survey; the procedure converts the camera-based MaxN into an absolute GAJ abundance estimate, alongside uncertainty estimates. As the dataset from our calibration experiment is very small, the model used to develop the conversion from MaxN to absolute abundance is substantially simpler than the general framework. To illustrate the potential utility of the general framework in data rich calibration studies, we present a simulation study in section~\ref{sec-sim}. In section~\ref{sec-adjust}, we propose an approach to handle the case in which both a camera and the echosounder are deployed simultaneously during the survey --- instead of modeling GAJ abundance, the approach in this case derives an adjustment factor for the acoustic count that is more robust than a factor derived based solely on the single camera type deployed. The procedures from Sections~\ref{sec-formulae} and~\ref{sec-adjust} are implemented as a single widget, which is described in these sections. Final thoughts and future work are discussed in Section~\ref{sec-tbd}.

\section{DATA FROM THE CALIBRATION EXPERIMENT}\label{sec-calibdata}

Figure~\ref{fig-map} shows the locations of all 13 unique reef structures (east of New Orleans, Louisiana and south of Mobile, Alabama) visited in the calibration experiment. Some columns of the data table appear in Figure~\ref{fig-df}.

\begin{figure}
\centering{
\includegraphics[width=.85\linewidth]{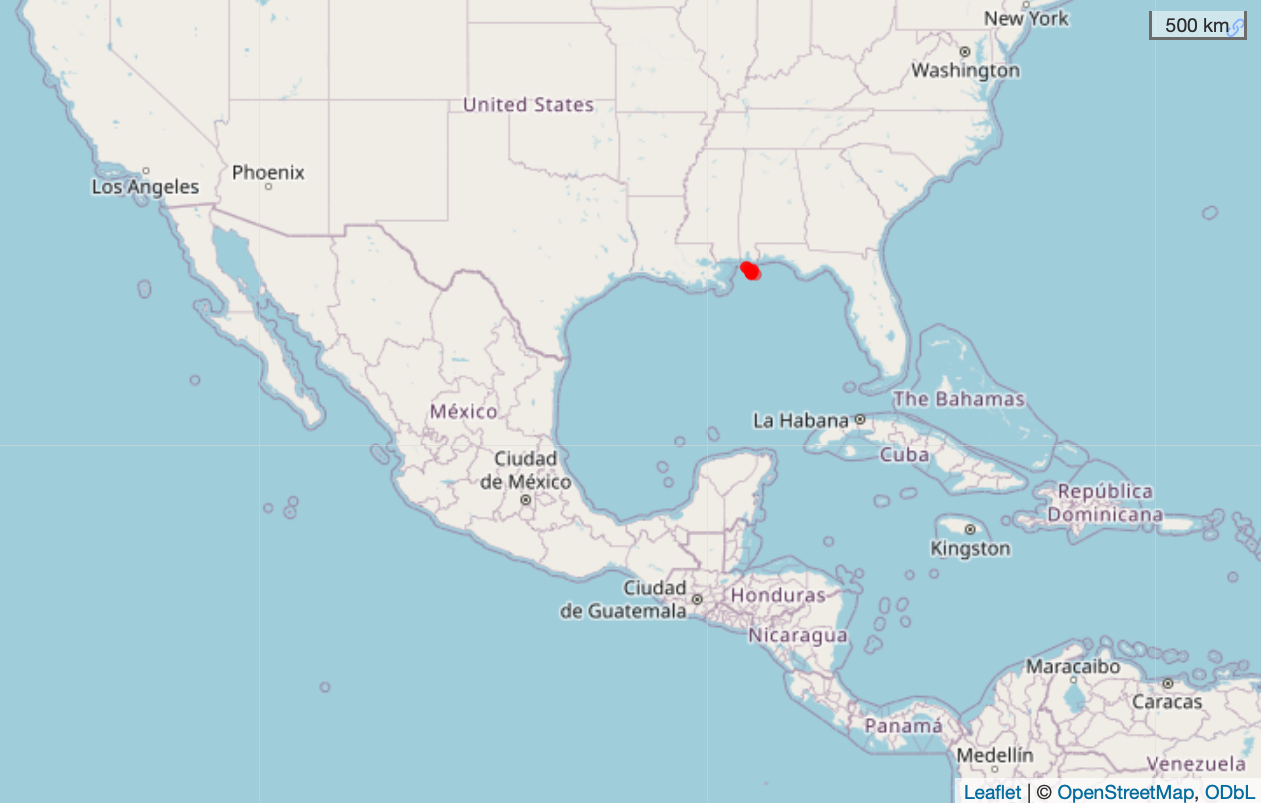}

\includegraphics[width=.85\linewidth]{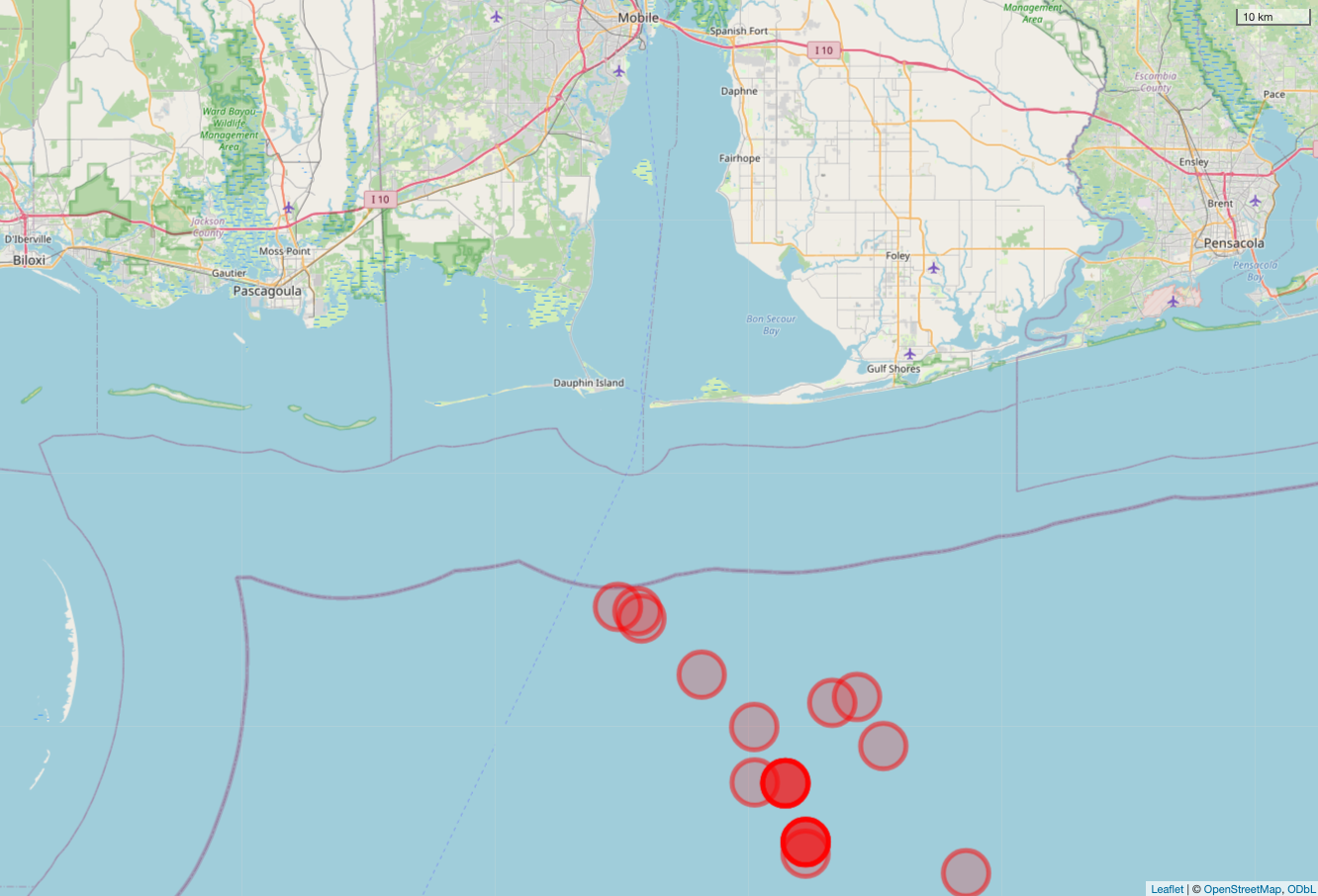}
}
\caption{\label{fig-map} Maps showing the site of the calibration experiment, displayed at two spatial scales: zoomed out (top panel) and zoomed in (bottom panel). Red markers denote reef structures visited in the experiment. The two darker markers in the bottom panel denote multiple visits to the same structure (5 trips each).}
\end{figure}

\begin{figure}
\centering{
\includegraphics[scale=.5]{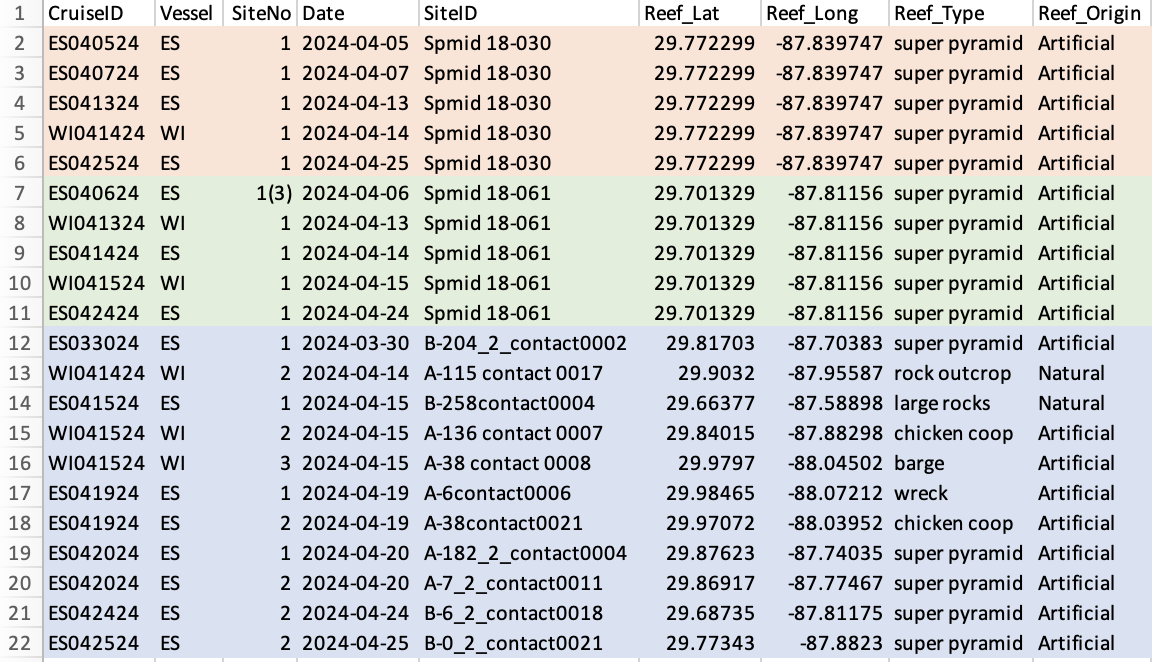}
}
\caption{\label{fig-df}Some identifying features of the 21 boat trips in
the calibration experiment.}
\end{figure}

A total of 13 unique reef structures were visited by 21 separate boat
trips; different structures visited by the same boat on the same day
constitute separate trips; similarly for the same structure visited on
different dates, e.g., the artificial super pyramid structure
\texttt{Spmid\ 18-030} generated five trips, and similarly for the
artificial super pyramid \texttt{Spmid\ 18-061}.

Each \(s\)th trip is associated with key abundance-related variables
that include

\vspace{5mm}

\begin{itemize}
\item
\(y_{s\ell}=\) video-derived MaxN for GAJ based on the \(\ell\)th
camera type --- thus, \(y_{s\ell}\) is missing (denoted by \texttt{NA}) if type \(\ell\)
camera was missing; all 21 trips involved both SBRUV and trap
cameras, 2 trips lacked the drop camera only, 5 lacked the ROV camera
only, and 2 lacked both the drop and ROV cameras;
\item
\(N_s=\sum_t N_{st}\) (never \texttt{NA}) is the total count of GAJ and GAJ-like fish (collectively denoted by ``GAJ+'') derived
from acoustic signals along all transects combined, i.e., the sum of \(N_{st}\)
across all \(t\)th transects traversed by the echosounder;
\item
\(N_s^{(f)}=\) total acoustic-derived count of GAJ+ (never \texttt{NA}) pooled from focal transects only, i.e., transects that are most
relevant to GAJ activities around the $s$th reef structure --- thus, $N_s^{(f)} \le N_s$;
\item
\(N_s^{(mr)}=\) Lincoln-Petersen estimate \citep{Amstrup2005} of absolute GAJ abundance
based on standard mark-recapture protocol administered at the structures
\texttt{Spmid\ 18-030} and \texttt{Spmid\ 18-061} only (thus,
\(N_s^{(mr)}\) is \texttt{NA} for 11 trips, and actual observations are always at the same type of habitat, being a ``super pyramid''); and
\item
scaling factor
\[
r_s = \frac{\sum_{\{\ell:\ell\text{ present}\}} y_{s\ell}}{\sum_{\{\ell:\ell\text{ present}\}} \{(s,\ell)\text{th MaxN for all large-bodied species in the water column\}}} + 10^{-6}
\]
(see Appendix~\ref{sec-GAJ+} for the list of species
 included in the denominator of $r_s$) --- the shift factor of \(10^{-6}\) avoids log(0) in the
calibration model.\\
\end{itemize}


In addition, each \(s\)th trip is associated with variables of a
potentially explanatory nature to abundance, including

\vspace{5mm}

\begin{itemize}
\item
reef origin --- of the 13 structures, 2 are natural and the rest are
artificial;
\item
reef type --- 6 types (see Figure~\ref{fig-df}), of which 3 (rock outcrop, natural; large rocks, natural; and chicken coop, artificial)
visited over 4 trips are considered small structures, and the other 3
types visited over 17 trips are considered large structures;
\item
boat ID --- 2 boats, one of which never carried an ROV camera; and
\item
date-boat combination --- 14 combinations across 21 trips.\\
\end{itemize}

The above variable descriptions reveal that the calibration experiment
exhibited an extremely unbalanced experimental design with very few data
points. This lopsidedness leads to substantial modeling challenges:
  missing data due to \texttt{NA} under many treatment combinations would
cause many explanatory effects to be inestimable. To preemptively
mitigate inestimability challenges, we assume that

\vspace{5mm}

\begin{itemize}
\item
reef size (large or small under reef type) encapsulates any effects on
abundance due to reef origin and reef type, and
\item
trip date is irrelevant to abundance.\\
\end{itemize}

Thus, our unifying model can only include a maximum of two explanatory
variables (boat and reef size) and possibly their interaction with
camera type. Even with a reduced set of candidate explanatory variables,
severe imbalance remains in the experimental design, e.g., \(\ell=R\)
appeared on only one of the boats, and only 4 of 21 trips involved small
reef structures. The imbalance is exacerbated when \(N_s^{(mr)}\) is
included in the model for exploration/validation purposes, as
mark-recapture data are associated with only two large structures over
10 trips; this results in painstaking adjustments being required each
time a potential explanatory variable is omitted from or added to the
model (see Appendix~\ref{sec-appdx-timeline} for details).

\subsection{VISUALIZING THE EXPERIMENTAL DATA}\label{visualizing-the-experimental-data}

\begin{figure}
\centering{
\includegraphics[scale=.2]{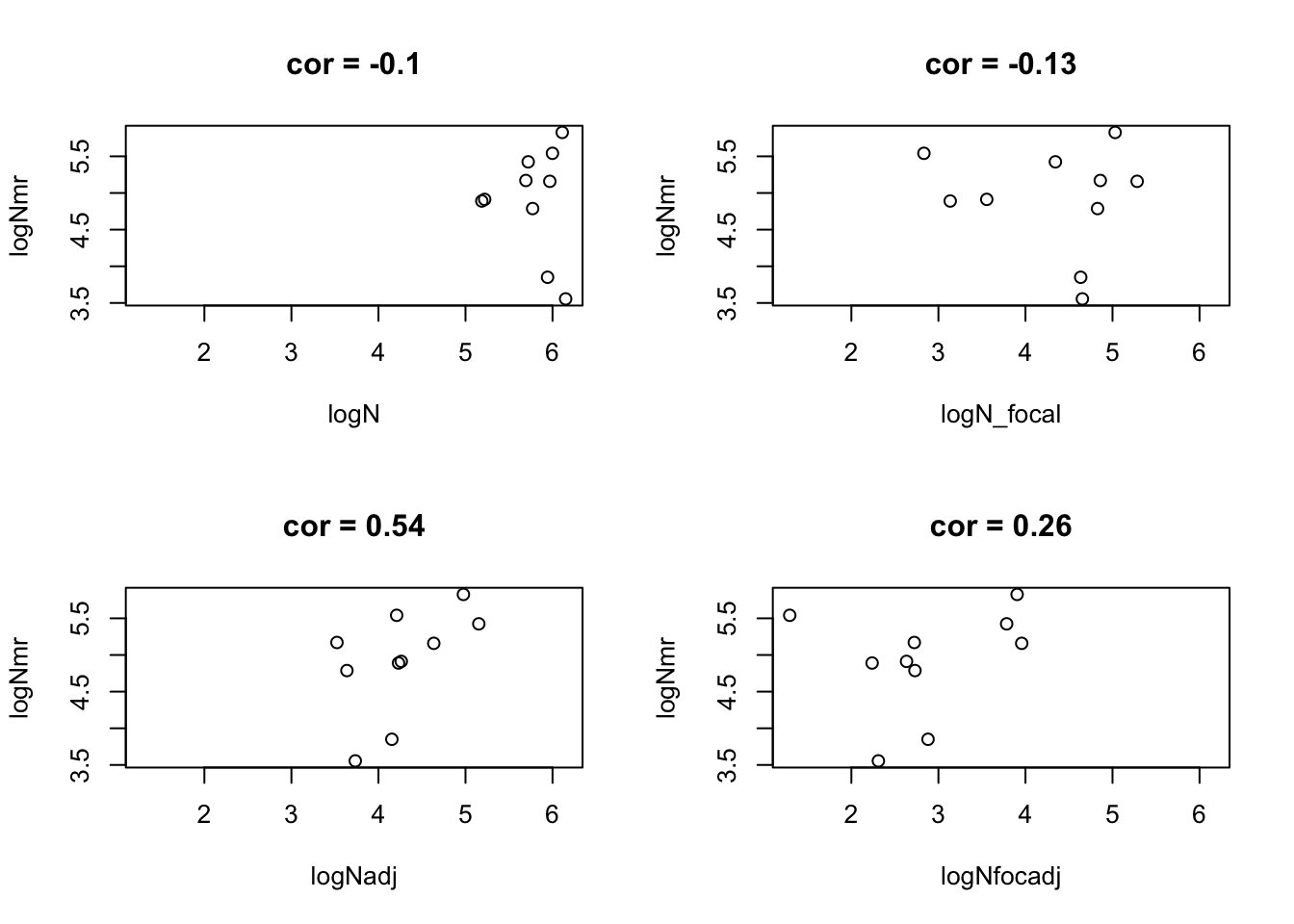}
}
\caption{\label{fig-NvsNmr}\textbf{Mark-recapture abundance estimates
(\(N_s^{(mr)}\)) vs.~acoustic counts (\(N_s\) or \(N_s^{(f)}\)),} the
latter being unscaled (top row) or scaled by the camera ratio \(r_s\)
pooled across all four camera types (bottom row). All axes are on the
log scale after adding 1 to avoid log(0).}
\end{figure}

Species count data are best visualized on the log scale due to
substantially skewed distributions.

As can be seen from the top panels in Figure~\ref{fig-NvsNmr}, unscaled
acoustic counts (\(N_s\) or \(N_s^{(f)}\)) are somewhat negatively
correlated with mark-recapture estimates, but the bottom panels show
that the correlation is noticeably positive when the acoustic counts
have been scaled by the pooled camera ratios (\(r_s\)). Similarly, from
the top panels of both Figures~\ref{fig-cam-vs-N} and~\ref{fig-cam-vs-Nfoc}, we see that MaxN counts (\(y_{s\ell}\))
and unscaled acoustic counts (\(N_s\) or \(N_s^{(f)}\)) generally do not
correlate if the smallest counts (all of which originate from small reef
structures) are omitted; the bottom panels show that the correlation
improves noticeably when the acoustic counts have been scaled. All of
Figures~\ref{fig-NvsNmr}, \ref{fig-cam-vs-N}, and~\ref{fig-cam-vs-Nfoc} indicate that focal acoustic counts
(\(N_s^{(f)}\)) have weaker correlations with MaxN counts and
mark-recapture estimates when compared to acoustic counts pooled across
all transects (\(N_s\)).

\begin{figure}
\centering{
\includegraphics[scale=.2]{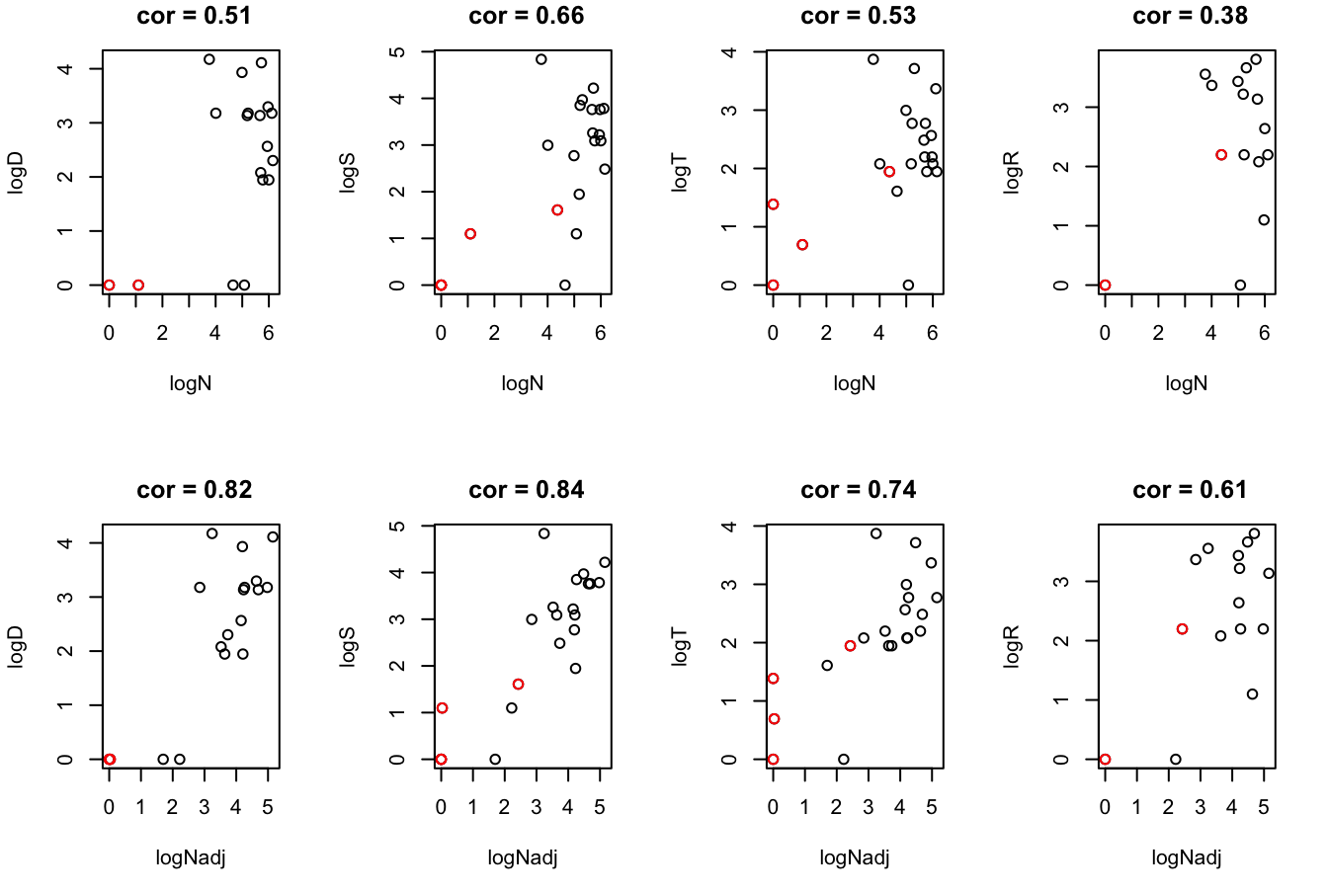}
}
\caption{\label{fig-cam-vs-N}\textbf{Camera-specific MaxN
(\(y_{s\ell}\)) vs.~acoustic count (\(N_s\))}, the latter being unscaled
(top row) or scaled by \(r_s\) (bottom row). Both axes are on the log
scale after adding 1 to avoid log(0). Red circles denote data points at
small reef structures. Each correlation coefficient accounts for both
small and large reefs.}
\end{figure}

\begin{figure}
\centering{
\includegraphics[scale=.2]{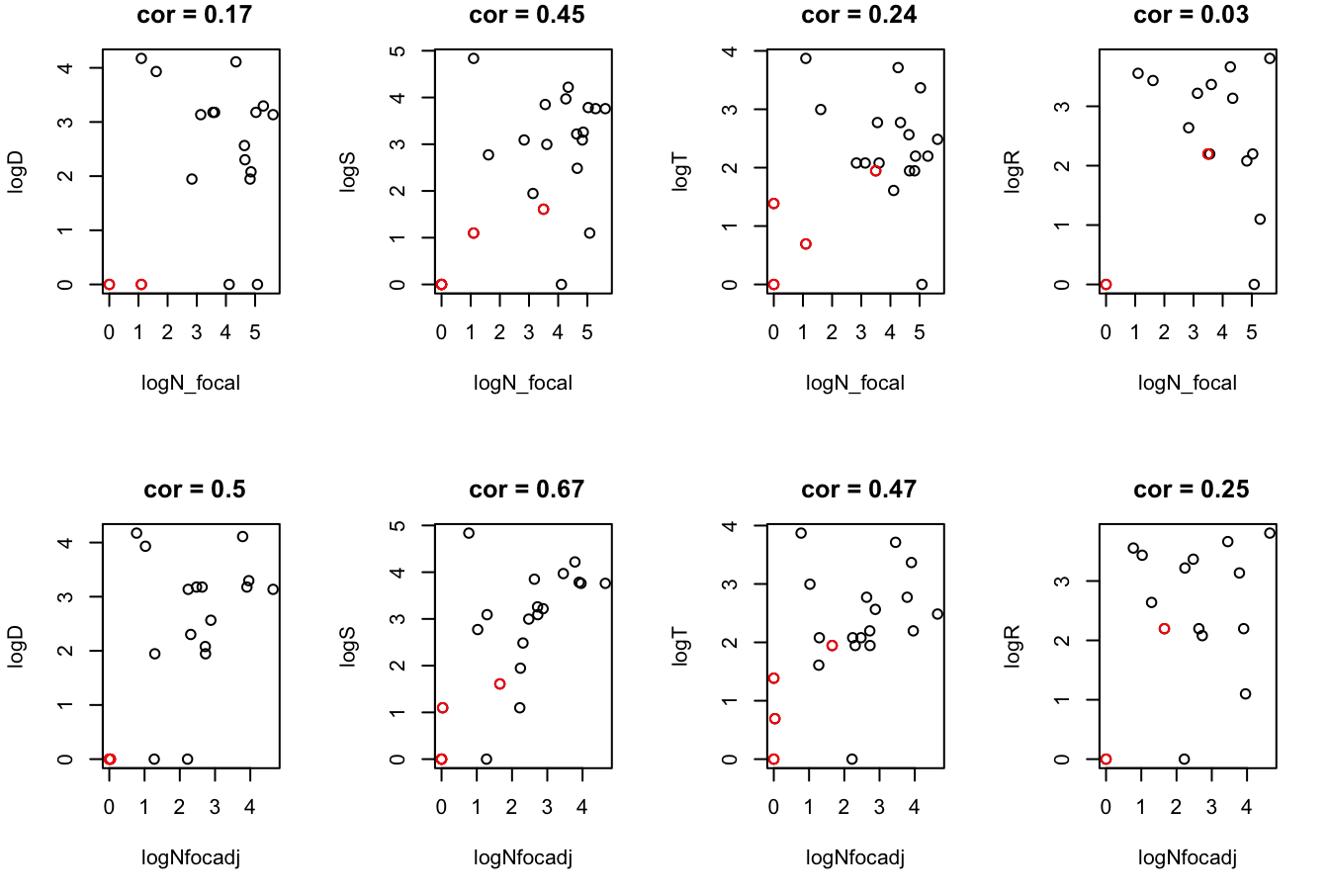}
}
\caption{\label{fig-cam-vs-Nfoc}\textbf{Camera-specific MaxN
(\(y_{s\ell}\)) vs.~focal acoustic count (\(N_s^{(f)}\))}. See
Figure~\ref{fig-cam-vs-N} caption for additional details.}
\end{figure}

With respect to camera ratios, Figure~\ref{fig-camratios} indicates that
camera-specific ratios are weakly correlated among the four camera
types. In contrast, the pooled ratio, \(r_s\), correlates reasonably
well with each of the camera-specific ratios, providing justification
for our pooling method.

\begin{figure}
\centering{
\includegraphics[scale=.2]{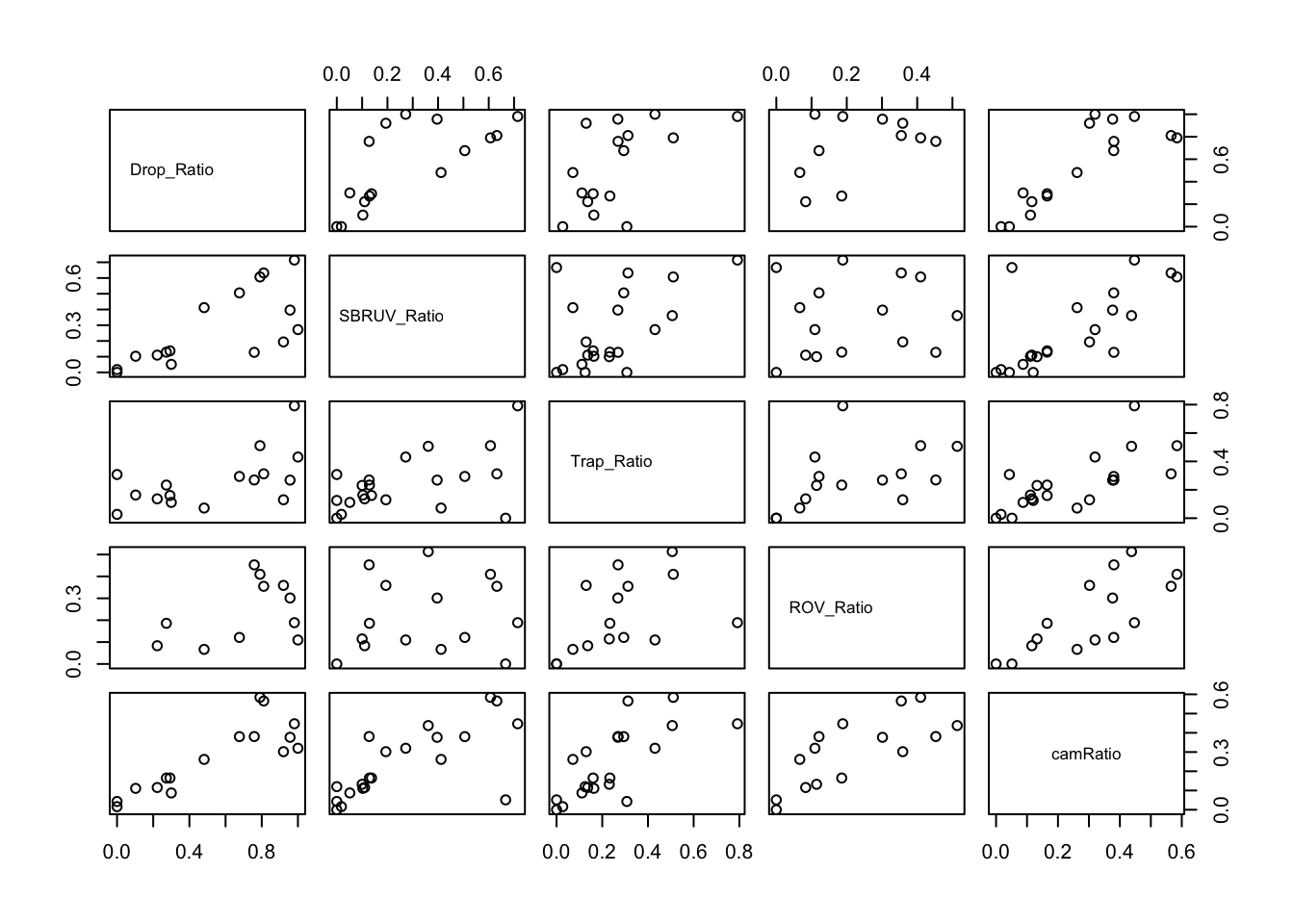}
}

\caption{\label{fig-camratios}\textbf{Scatter-plot matrix among the
pooled ratio \(r_s\) (labeled as \texttt{camRatio}) and camera-specific ratios}
(\(=y_{s\ell}/\{(s,\ell)\text{th MaxN for all GAJ+}\}\),
unpooled).}
\end{figure}

\section{CALIBRATION MODELING FRAMEWORK}\label{sec-framework}

If the entire field survey involves MaxN's from a single camera type (paired with acoustic data or otherwise), then a traditional calibration approach may be to run a calibration experiment that pairs the camera with the echosounder, fit a simple regression of acoustic count
on MaxN to these data, then use the regression
curve as the calibration formula to convert any future MaxN from the same camera type to an absolute
abundance estimate; the confidence bands around the
regression curve would provide an estimate of the calibration
uncertainty. However, in our study, (i) a MaxN reflects the
species of interest but an acoustic count reflects multiple species combined, and (ii)
each region in the continental-scale field survey deploys its own camera type.

Thus, our key objective is to construct a unifying statistical modeling
framework that reflects the single, joint relationship among all
camera-specific MaxN counts of GAJ (\(y_{Ds}, y_{Ss}, y_{Ts},\) and \(y_{Rs}\)) and the
acoustic count of GAJ+ (either \(N_s\) or \(N_s^{(f)}\)). Such a cross-calibration
modeling framework replaces a series of simple regression fits; we construct a
single Bayesian hierarchical model that jointly relates the vector \(\boldsymbol{y}_s\) to \(N_s\) or \(N_s^{(f)}\) observed in the calibration experiment. The framework allows boat and reef size as predictor variables, while
the pooled camera ratio \(r_s\) enters the model as the offset term in
the Poisson log-link function for the true mean GAJ abundance.

The calibration model under this framework is trained by data from the
21 boat trips in the calibration experiment discussed in
Section~\ref{sec-calibdata}, in which all trips involved the acoustic
gear and many trips involved all four camera types.

\subsection{KEY ASSUMPTIONS}\label{key-assumptions}

Our methodology assumes that

\vspace{5mm}

\begin{itemize}
\item
the scaling factor $r_s$
is a reasonable approximation of the true ratio between the number of
GAJ individuals and the number of all acoustically indistinguishable
individuals,
\item
our calibration experiment --- involving 13 reef structures visited by
a total of 21 boat trips combined --- provides a reasonable basis for
training the joint relationship between
\(\boldsymbol{y}_s=[y_{sD}, y_{sS}, y_{sR},y_{sT}]'\) and \(N_s\) or \(N_s^{(f)}\)
  across the study-wide region of interest, so that
\item
  regional absolute abundance estimates, computed based on the
  calibration modeling methodology described in this paper, are
  justifiably additive, i.e., can be totaled into a study-wide estimate
  of GAJ absolute abundance.\\
\end{itemize}


\subsubsection{Caveats}\label{caveats}

We acknowledge that MaxN counts are not necessarily additive, so that
\(r_s\) as defined above may be an unreliable approximation of the ratio
between GAJ count and GAJ+ count, although an improved
definition of \(r_s\) would require video processing that is deemed
impractical given the scope of this project. Notwithstanding, cross-validation against mark-recapture abundance estimates suggests that $r_s$ as defined is a reasonable correction factor for acoustic data (see Section \ref{sec-final-model}).  

Secondly, although we use a single list of species deemed to be
``large-bodied'' across the study (see Appendix~\ref{sec-GAJ+}), the MaxN counts for any
given ``large-bodied'' species may include individual fish that is
smaller in size than those recognized by the algorithm that converts
echosounder signals to a fish count.

Thirdly, the calibration experiment was confined to a tiny, isolated
swath inside the study-wide spatial domain, with possibly limited
representation of the regional conditions and habitat features across
the study; however, challenges of this nature were unavoidable due to
the lack of existing pilot calibration studies prior to the current
continental-scale project with very ambitious objectives.

Finally, neither \(N_s\) nor \(N_s^{(f)}\) is the ideal acoustic data
for calibration purposes. Instead, a value (e.g., from interpolation)
that reflects the type of structure (i.e., point-level or areal) would
be necessary for all resulting acoustic counts to share a common
areal/voluminal denominator (see Appendix~\ref{sec-appdx-acoustic}), although such an undertaking for all data from the full-scale study will require ongoing data processing beyond the scope of this paper.

\subsection{MOST COMPREHENSIVE FRAMEWORK}\label{sec-complex-model}

Our most comprehensive --- hence, ambitious --- Bayesian hierarchical
mixed-effects framework jointly models camera-specific MaxN counts
(\(y_{s\ell}\) where \(\ell=D,S,T,R\)), acoustic counts (\(N_s\)), and
mark-recapture estimates (\(N_s^{(mr)}\)), while accounting for boat
(two types), reef size (small or large), and their possible interactions
with camera type (four types). With only 21 boat trips, each of which
has at least one missing value for \(y_{s\ell}\) and \(N_s^{(mr)}\) due
to missing \{boat, reef, camera type\} combinations, the model is
expected to be inestimable unless upfront constraints are stipulated.
Thus, our framework upfront disregards the interaction between boat and
reef size, but it allows the interaction between camera type with each
of boat and reef size. It also allows reef-size-specific regression
error variances within the model hierarchy. Slope parameters are
constrained to be non-negative, and fixed effects must sum to zero. The
count variables, namely, \(y_{s\ell}, N_s,\) and \(N_s^{(mr)}\), are all
modeled as conditional Poisson random variables, whose conditional means
are log-normally distributed random quantities. The four MaxN counts are
explicitly expressed as multivariate in the model hierarchy.

Instead of indexing the boat trips by \(s=\) 1, 2, \ldots, 21, we
reindex the boat trips as boat-reef-specific: trip \(k_{ij}\) is the
\(k\)th trip for the \((i,j)\)th boat-reef combination, where
\(i,j=1,2\), and \(k_{11}=1,2,...,13\); \(k_{12}, k_{22}=1,2\); and
\(k_{21}=1,2,3,4\).

Next, let \(\mu\) denote expected MaxN counts, \(\tau\) denote expected
acoustic and mark-recapture counts, \(\phi\) denote the underlying true
expected GAJ abundance, \(\beta_0\) and \(\beta_1\) denote regression
intercept and slope parameters, respectively; \(\nu, \gamma\), and
\(\xi\) denote fixed effects for boat, reef size, and absolute abundance
count type, respectively; \(\varepsilon\) and \(\zeta\) denote linear
regression noise terms, \(\sigma^2\) denote regression error variances,
and \(\rho\) (constrained to be non-negative) denote the inter-MaxN
correlation in an exchangeable correlation structure; interactions are
denoted by index combinations. Note that all \(\mu\)'s and \(\tau\)'s
are modeled as functions of \(\phi\); thus, the underlying true expected GAJ
abundance, \(\phi\), is the single model parameter that unifies the
relationship among all MaxN counts, acoustic counts, and mark-recapture
estimates. Then, our comprehensive hierarchical model is

\begin{align*}
\text{Level 1.1: \quad} && y_{ijk\ell} &\sim \text{Pois}(\mu_{ijk\ell}) \\
\text{Level 1.2: \quad} & \text{for } h=1 \text{ (acoustic)}: & N_{ijk} &\sim \text{Pois}(\tau_{hijk}) \quad \text{for all } i,j,k \\
& \text{for } h=2 \text{ (mark-recapture)}: & N_{i1k}^{(mr)} &\sim \text{Pois}(\tau_{hi1k}) \quad \text{for all } i,k \quad (j=1 \text{ only}) \\
\text{Level 2.1: \quad} & \text{for } \ell=D,S,T: & \log\mu_{ijk\ell} &= \beta_{y,0,\ell} + \nu_{y,i\ell} + \gamma_{y,j\ell} + \beta_{1,j\ell}\widetilde{\log\phi}_{ijk} + \varepsilon_{ijk\ell} \\
&& \left[ \begin{matrix}
\varepsilon_{ijkD} \\
\varepsilon_{ijkS} \\
\varepsilon_{ijkT} 
\end{matrix} \right] &\sim \text{MVN}(\mathbf{0}, \sigma_y^2\mathbb{P}) \\ 
&& \mathbb{P}_{\ell_1,\ell_2} = \mathbb{P}_{\ell_2,\ell_1} &= \left\{
  \begin{matrix}
  \rho & \text{if } \ell_1\ne\ell_2 \\
  1 & \text{if } \ell_1=\ell_2
  \end{matrix} \right. \\
& \text{for } \ell=R: & \log\mu_{ijkR} &= \beta_{y,0,R} + \gamma_{y,jR} + \beta_{1,jR}\widetilde{\log\phi}_{ijk} + \nonumber \\
&& &\quad\quad \beta_{1,D}\widetilde{\log\mu}_{ijkD} + \beta_{1,S}\widetilde{\log\mu}_{ijkS} + \nonumber\\
&& &\quad\quad\quad \beta_{1,T}\widetilde{\log\mu}_{ijkT} + \varepsilon_{ijkR} \\
&& \varepsilon_{ijkR} &\sim \mathcal{N}(0, \sigma_{yR}^2) \\
& \text{for all } \ell: & 0 = \sum_i \nu_{y,i\ell} &= \sum_j \gamma_{y,j\ell}\quad \text{for identifiability} \\
\text{Level 2.2: \quad} & \text{for } h=1: & \log(r_{ijk}\tau_{hijk}) &= \log\phi_{ijk} + \xi_1 + \zeta_{hijk} \\
& \text{for } h=2: & \log\tau_{hi1k} &= \log\phi_{i1k} + \xi_2 + \zeta_{hi1k} \quad (j=1 \text{ only}) \\
& \text{for } h=1,2: & \zeta_{hi1k} &\sim \mathcal{N}(0, \sigma_{x,1}^2) \\
& \text{for } h=1,2: & \zeta_{hi2k} &= 0 \quad \text{for identifiability due to } k_{i2}\le 2 \\
&& 0 &= \sum_h \xi_h \quad \text{for identifiability} \\
\text{Level 3: \quad} && \log\phi_{ijk} &= \beta_0 + \nu_{x,i} + \gamma_{x,j} + \delta_{ijk} \\
&& \delta_{ijk} &\sim \mathcal{N}(0,\sigma_{\phi,j}^2) \\
&& 0 = \sum_i \nu_{x,i} &= \sum_j \gamma_{x,j} \quad \text{for identifiability}
\end{align*}
where, for computational stability, $\widetilde{\log\mu}_{ijk\ell}$ is taken to be either $\log\mu_{ijk\ell}$ (uncentered) or $\log\mu_{ijk\ell}-(1/21)\sum_{i,j,k}\log\mu_{ijk\ell}$ (centered), and similarly for $\widetilde{\log\phi}_{ijk}$. We employ relatively diffuse prior distributions
\begin{align}
\text{logit}\, \rho &\sim \mathcal{N}(0,2^2) \label{prior1} \\
\log\sigma &\sim \mathcal{N}(0,2^2)\ \text{ for all}\ \sigma\text{'s} \\
\log\beta_1 &\sim \mathcal{N}(0, 2^2) \text{ or } \mathcal{N}(0, 3^2)\ \text{ for all}\ \beta_1\text{'s} \\
\beta_0, \gamma, \nu, \xi &\sim \mathcal{N}(0,3^2)\ \text{ for all}\ \gamma\text{'s}, \nu\text{'s} \label{priorLast}
\end{align}

\vspace{5mm}

Note that in Level 2.1 of the hierarchy, the quadvariate relationship is
broken down into a trivariate normal for drop-, SBRUV-, and trap-based
MaxN regression noise terms, and a univariate log-linear regression of
the mean ROV-based MaxN on the other three log-transformed MaxN means
and reef size effect. This breakdown reflects the fact that ROV cameras
were never deployed on one of the two boats.

The comprehensive model of this subsection has the following features:

\vspace{5mm}

\begin{itemize}
\item
  The true expected abundance (\(\phi\)) can depend on boat (\(\nu\)) and
  reef size (\(\gamma\)), and the regression error variance
  (\(\sigma_{\phi}^2\)) can be reef-specific (i.e., indexed by \(j\)).
\item
  On the log-mean scale, both the expected acoustic and expected
  mark-recapture counts (\(\tau\))

  \begin{itemize}
  \item
    are equal to the true expected abundance (\(\phi\)), but possibly
    subject to a count-type-specific offset (\(\xi\)) and noise
    (\(\zeta\)); and
  \item
    have a regression error variance (\(\sigma_x^2\)) that can be
    reef-size-specific (i.e., indexed by \(j\)).\footnote{Note the
      constraint \(\sigma_{x,j=2}^2=0\) for expected 
      counts at small reefs (model Level 2.2), which is equivalent to
      \(\zeta_{j=2}=0\).}
  \end{itemize}
\item
  On the log-mean scale, the expected value of MaxN counts (\(\mu\)) that are observed at both
  reef sizes (\(\ell=D,S,T\)) can depend on the true expected abundance
  (\(\phi\)), boat (\(\nu\)), and reef size (\(\gamma\)). Moreover,

  \begin{itemize}
  \item
    any dependence on boat and reef size at the MaxN-level is in
    addition to the dependence on boat and reef size at the
    \(\phi\)-level;
  \item
    the effect of \(\phi\) on \(\mu\) (i.e., slope parameter
    \(\beta_1\)) can be reef-size-specific (i.e., indexed by \(j\));
    and
  \item
    the regression noise terms (\(\varepsilon\)) follow a multivariate
    normal distribution with mean 0 and exchangeable correlation
    structure (i.e., all pairwise correlations are \(\rho\)) with a
    common error variance (\(\sigma_y^2\)) that is not specific to boat
    or reef size.
  \end{itemize}
\item
  On the log-mean scale, the expected value of ROV MaxN counts that are observed on one boat only
  (\(\mu_{\ell=R}\))

  \begin{itemize}
  \item
    can depend on the other expected MaxN counts (\(\mu_{\ell\ne R}\))
    and reef size (\(\gamma\)), but not on boat (\(\nu\));
  \item
    the effect of \(\phi\) on \(\mu_{\ell=R}\) (i.e., slope parameter
    \(\beta_{1R}\)) can be reef-size-specific (i.e., indexed by
    \(j\));
  \item
    the effects of other expected MaxN counts (\(\mu_{\ell\ne R}\)) on
    \(\mu_{\ell=R}\) (i.e., slope parameters
    \(\beta_{1D},\beta_{1S},\beta_{1T}\)) are constant across reef
    sizes; and
  \item
    the regression noise term (\(\varepsilon_{\ell=R}\)) does not share
    the same error variance (\(\sigma_y^2\)) as those for the other
    expected MaxN counts (\(\varepsilon_{\ell\ne R}\)).\\
  \end{itemize}
\end{itemize}

While the complexity of the comprehensive framework presented above may be impractical for the small dataset from our calibration experiment, we present a simulation study in Section~\ref{sec-sim} that illustrates the potential utility of our calibration model in scenarios that are more data rich.

\subsection{SIMPLER MODELS}\label{sec-other-models}

For the 21 observations from our calibration experiment, we consider simpler versions of the general model in Section~\ref{sec-complex-model}. They include a combination of the following
model reductions:

\vspace{5mm}

\begin{itemize}
\item
  Omit \(h=2\) from Level 1.2, thus collapsing Levels 2.2 and 3 into a
  single level and omitting any inestimable quantities that pertain to
  the presence of two sets of absolute abundance data.
\item
  Omit \(r\) from consideration, thus, \(\phi\) is modeled as the
  expectation of the raw acoustic count \(N\) in the absence of \(h=2\),
  and of \(N\) with an offset \(\xi\) in the presence of \(h=2\).
\item
  Omit \(\nu\)'s in Level 2.1, thus collapsing the \(\ell=R\) and
  \(\ell\ne R\) cases into a single log-linear regression of \(\mu\) on
  \(\phi\) and reef size effect, with a quadvariate normal distribution
  for the regression error terms.
\item
  Omit any of $\{\beta_{y,0}\}, \{\nu_x\}, \{\nu_y\}, \{\gamma_x\}$, and
  $\{\gamma_y\}$.
\item
  Taking \(\beta_{1,j\ell_1}=\beta_{1,j\ell_2}\) for some
  \(\ell_1,\ell_2\).\\
\end{itemize}

Each model version considered (see
Appendix~\ref{sec-appdx-timeline}) is implemented using the NIMBLE Bayesian
programming software for R \citep{Ponisio2020, nimble-software:2024}, which produces Markov chain Monte Carlo (MCMC) simulated draws
from the posterior distribution.

We consider a model to be adequate if all of the key criteria below are
met:

\vspace{5mm}

\begin{itemize}
\item
  Apparent convergence of the model, based on visual inspection of MCMC
  trace plots.
\item
  The posterior distribution suggests that all fixed-effects and slope
  parameters are credibly non-zero, and that \(\rho\) is credibly
  bounded away from 0 and 1.
\item
  The posterior distribution suggests a credibly close alignment between
  \(\phi\) (true expected GAJ abundance) and

  \begin{itemize}
  \item
    the camera-ratio-adjusted acoustic counts (i.e.,
    \(\phi_{ijk} \approx r_{ijk}N_{ijk}\) with high credibility) if
    \(r_{ijk}\) is part of the model; or
  \item
    the raw acoustic counts (i.e., \(\phi_{ijk} \approx N_{ijk}\) with
    high credibility) if \(r_{ijk}\) is omitted from the model.
  \end{itemize}
\item
  Cross-validation against mark-recapture estimates: the posterior distribution suggests a credibly positive correlation
  between \(\phi\) and $N^{(mr)}$ --- as $N^{(mr)}$ is an abundance estimate that is only available at one type of habitat (``super pyramid'') and thus may be inherently biased, we do not require $\phi\approx N^{(mr)}$ with high credibility so as to declare model adequacy.
\item
  The posterior distribution is noticeably different from the prior
  distribution, indicating that non-trivial Bayesian updating has taken
  place. \\
\end{itemize}

Many versions of our modeling framework exhibit poor model performance (see
Appendix~\ref{sec-appdx-timeline}). For
example, several exploratory versions that omit the camera ratios \(r\)
from the model yield very poor alignment between \(\phi\) and \(N\) on
both the count- and log-scales. That is, \(\phi \approx c_0 + c_1 N\) where
\(c_0\) and \(c_1\) are credibly very different from 0 and 1,
respectively; and the same is true when \(\phi\) and \(N\) are
replaced with \(\log\phi\) and \(\log (N+1)\).

\subsection{FINAL MODEL ADOPTED FOR PRODUCING CALIBRATION FORMULAE}\label{sec-final-model}

Among the model versions we have implemented for these experimental data, the least
complex yet still adequate version is

\begin{align*}
\text{Level 1: \quad} && y_{ijk\ell} &\sim \text{Pois}(\mu_{ijk\ell}) \\
\text{Level 2.1: \quad} && N_{ijk} &\sim \text{Pois}(\phi_{ijk}/r_{ijk}) \\
\text{Level 2.2: \quad} && \log\mu_{ijk\ell} &= \beta_{y,0,\ell} + \beta_{1,j\ell}\widetilde{\log\phi}_{ijk} + \varepsilon_{ijk\ell} \\
&& \left[ \begin{matrix}
\varepsilon_{ijkD} \\
\varepsilon_{ijkS} \\
\varepsilon_{ijkT} \\
\varepsilon_{ijkR}
\end{matrix} \right] &\sim \text{MVN}(\mathbf{0}, \sigma_y^2\mathbb{P}) \\ 
&& \mathbb{P}_{\ell_1,\ell_2} &= 
\left\{
  \begin{matrix}
  \rho & \text{if } \ell_1\ne\ell_2 \\
  1 & \text{if } \ell_1=\ell_2
  \end{matrix} \right. \quad \text{ for all } \ell_1,\ell_2 = D,S,T,R \\
\text{Level 3: \quad} && \log\phi_{ijk} &= \nu_{x,i} + \gamma_{x,j} + \delta_{ijk} \\
&& \delta_{ijk} &\sim \mathcal{N}(0, \sigma_{\phi,j}^2) \\
&& 0 = \sum_i \nu_{x,i} &= \sum_j \gamma_{x,j} \quad \text{for identifiability}
\end{align*}
with $\widetilde{\log\phi}_{ijk}$ being centered and prior distributions as given in Equations~(\ref{prior1})--(\ref{priorLast}). Note that first fitting the model variant with respective intercepts $\beta_{y,0,\ell}+\nu_{y,i\ell}+\gamma_{y,j\ell}$ in Level 2.2 and $\beta_0$ in Level 3 suggests that $\nu_{y,i\ell},\gamma_{y,j\ell},$ and $\beta_0$ are unnecessary.

\begin{figure}
\centering{
\includegraphics[scale=.2]{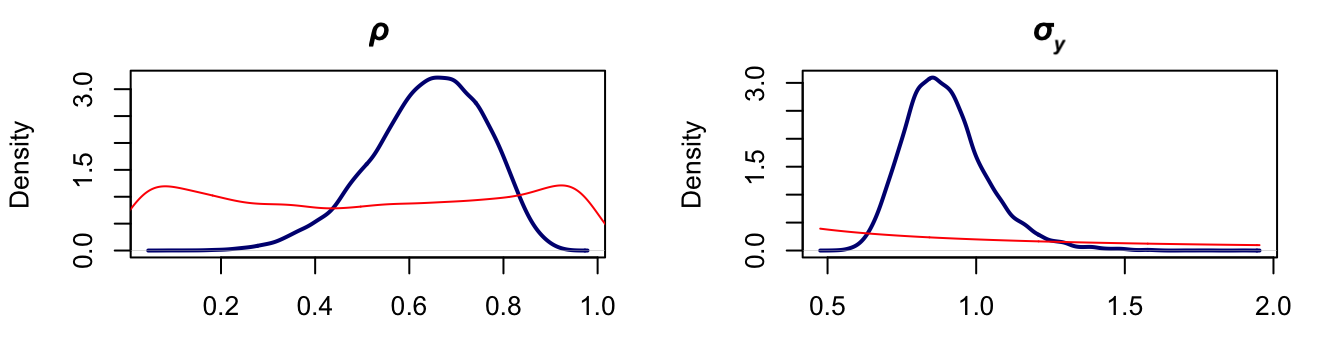}\\
\includegraphics[scale=.2]{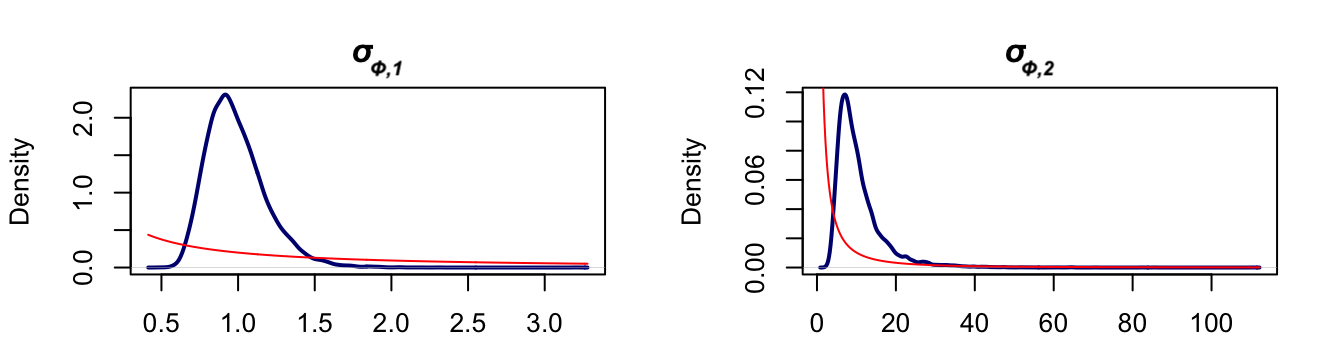}\\
\includegraphics[scale=.2]{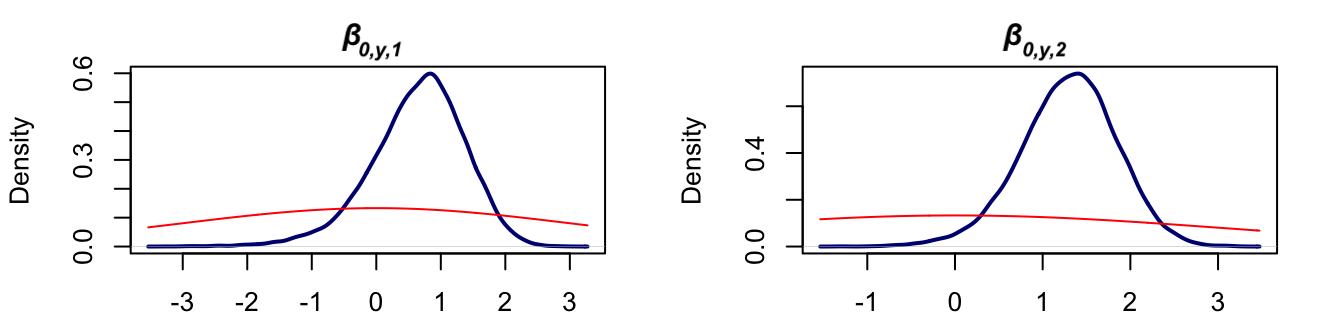}\\
\includegraphics[scale=.2]{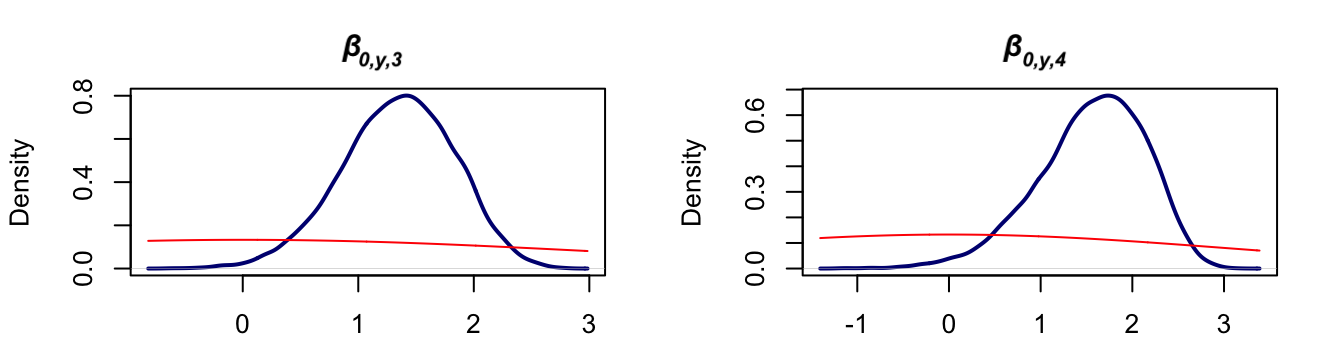}\\
\includegraphics[scale=.2]{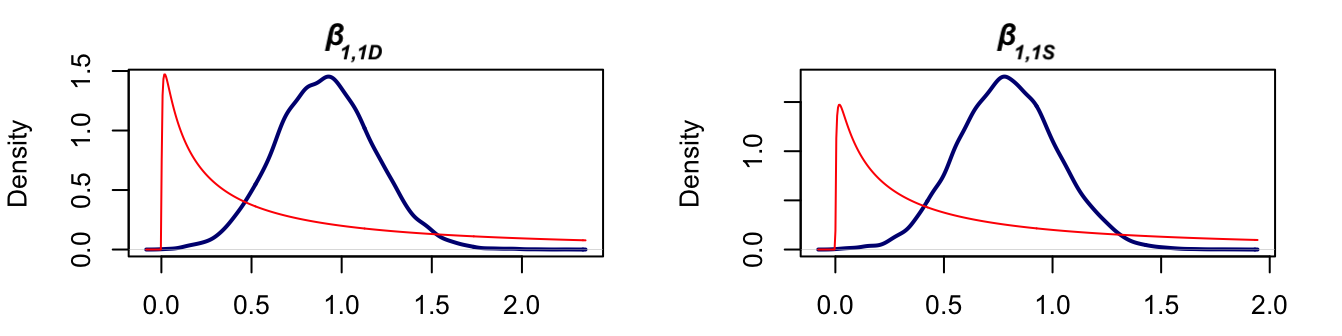}\\
\includegraphics[scale=.2]{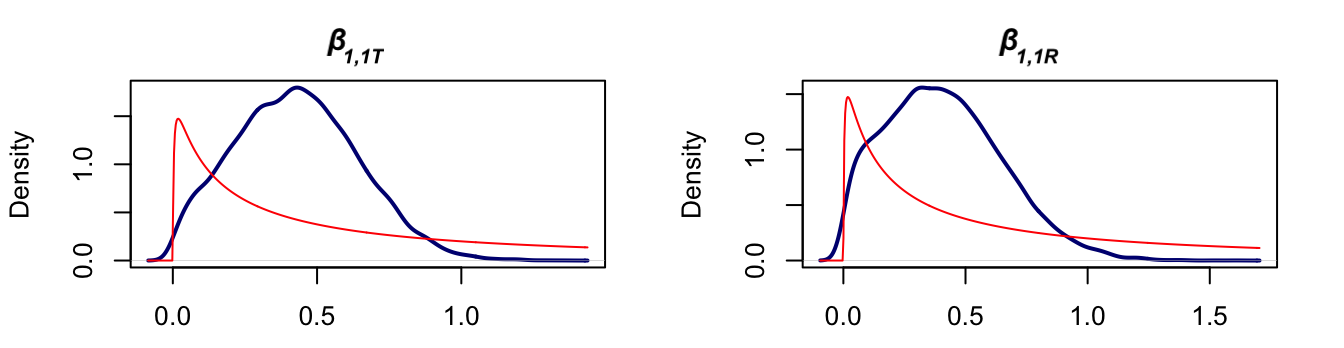}\\
\includegraphics[scale=.2]{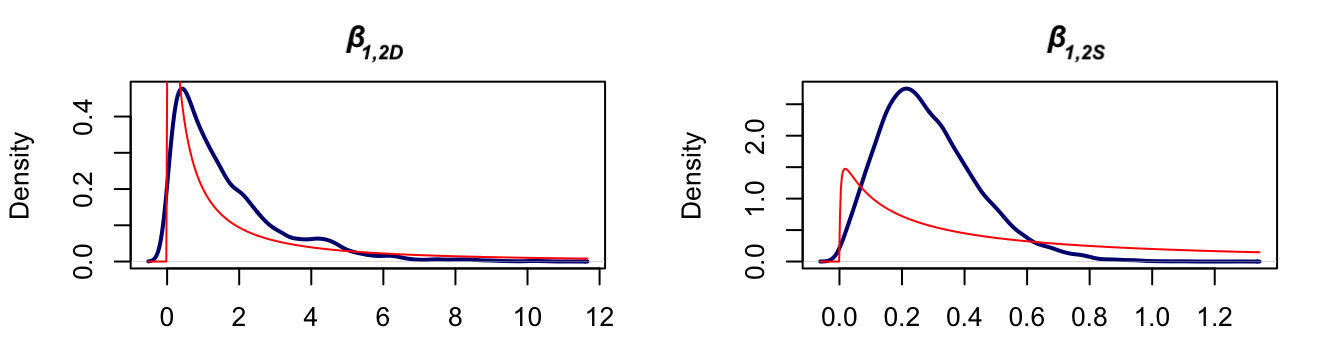}\\
\includegraphics[scale=.2]{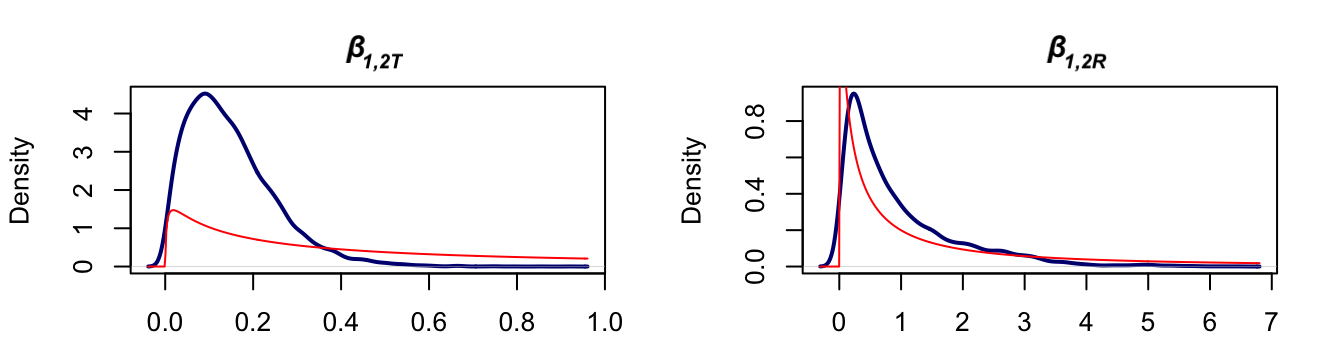}\\
\includegraphics[scale=.2]{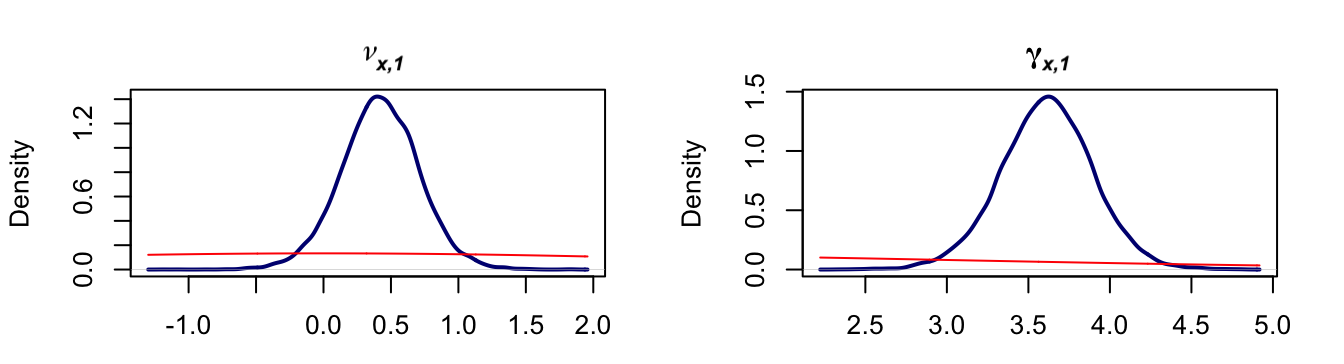}
}

\caption{\label{fig-post}\textbf{Posterior (blue) and prior (red) distribution} for all population-level model parameters.}
\end{figure}

Again, model adequacy is assessed according to the criteria given in Section~\ref{sec-other-models}. For example, Figures~\ref{fig-post} and \ref{fig-phi-vs-N} show non-trivial Bayesian updating and the desired alignment between $\phi$ and each of the mark-recapture count and the adjusted acoustic count. In particular, Figure~\ref{fig-post} shows that the marginal posterior distributions for population-level model parameters are noticeably different from the corresponding prior distributions, with the
unsurprising exceptions of \(\beta_{1,j=2,\ell=D}\) (left column, 7th panel) and \(\beta_{1,j=2,\ell=R}\) (right column, 8th panel) that correspond 
to some issues with drop and ROV camera data: at the four small reef structures, the pair $(y_{\ell=D}, y_{\ell=R})$ takes on the values (0, 0), (0, \texttt{NA}), (\texttt{NA}, 8), and (\texttt{NA}, \texttt{NA}), respectively, whereas $y_{\ell=S}$ and $y_{\ell=T}$ each takes on a variety of values across these four reefs. Figure~\ref{fig-phi-vs-N} indicates that, \textit{a posteriori}, the correlation between $\phi$ and $N^{(mr)}$ is high, and the regression line of $\phi$ on $rN$ has intercept $c_0\approx 0$ and slope $c_1\approx 1$. Note that Figure~\ref{fig-phi-vs-N} is produced as follows.

Upon fitting the final model, each \(M\)th draw from the posterior distribution is a vector
of all model parameter values, namely,
\(\{\mu_{ijk}^{(M)}, \phi_{ijk}^{(M)}, \beta_0^{(M)}, \beta_{1,j\ell}^{(M)}, \sigma_{\phi,j}^{(M)}, \sigma_y^{(M)}, \rho^{(M)}:\)
for all \(i,j,k,\ell\}\). Let us once again index a boat trip by \(s=\) 1, 2, \ldots, 21. Thus,
for each \(M\), frequentist least-squares regression can be applied to regress \(\{\phi_s^{(M)}: s=1,...,21\}\) on each of $\{N_s^{(mr)}: s=1,...,10\}$ and \(\{r_s N_s:s=1,...,21\}\) \textbf{while prohibiting a negative slope}; each $M$th regression line has intercept 
\(c_0^{(M)}\in (-\infty,\infty)\) and slope \(c_1^{(M)}\ge 0\). The median among the \(c_0^{(M)}\)'s and that among the \(c_1^{(M)}\)'s together yield the white ``median regression line'' in the panels of Figure~\ref{fig-phi-vs-N}.

\begin{figure}
\centering{
\includegraphics[scale=.2]{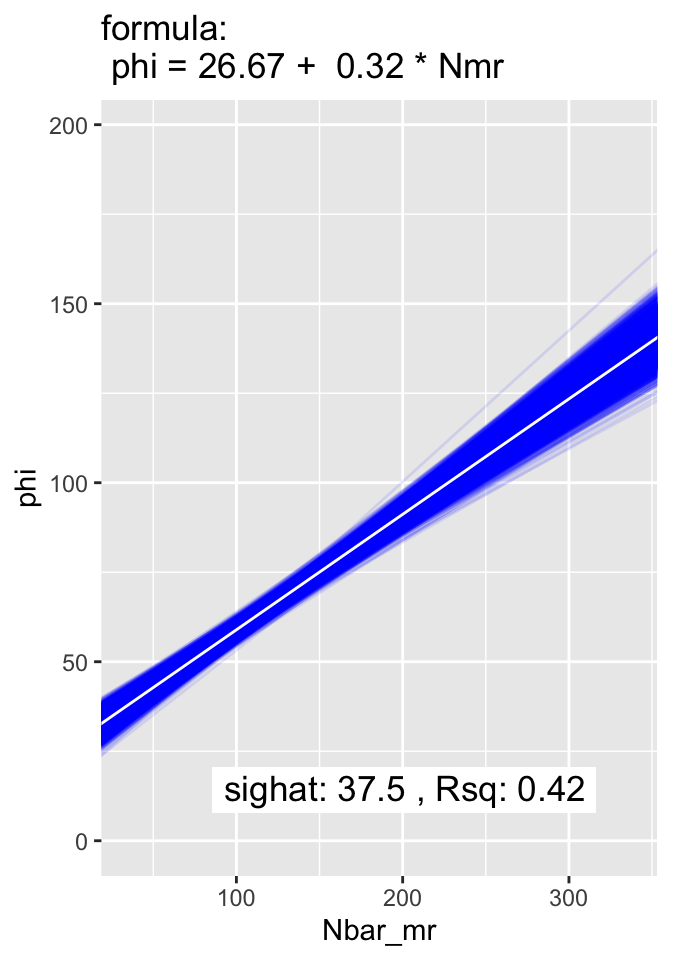} $\quad$
\includegraphics[scale=.2]{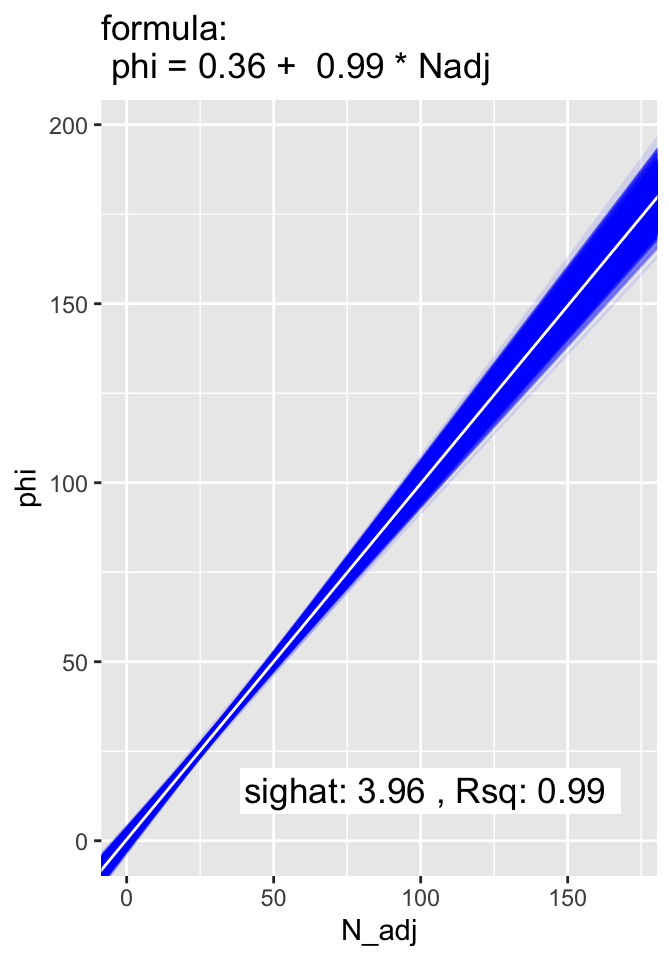} 
}

\caption{\label{fig-phi-vs-N}\textbf{Simple linear regression of modeled expected GAJ abundance on mark-recapture estimate (left) and adjusted acoustic count (right): 5000 least-squares regression lines (blue) and the ``median'' regression line (white).} Left panel: Each $M$th blue line is from taking $\{\phi_s^{(M)}: s=1,...,21\}$ --- the $M$th posterior draw for the vector $\boldsymbol{\phi}$ according to the final model --- and regressing it on $\{N_s^{(mr)}:s=1,...,10\}$ where mark-recapture data are available (\texttt{Nmr} or \texttt{Nbar\_mr} in the plot); displayed values for the intercept, slope, residual error SD (\texttt{sighat}) and $R^2$ are medians across the blue lines; the white line corresponds to the median intercept and slope displayed above the plot. Right panel: Same as the left panel, except regression on $\{r_s N_s: s=1,...,21\}$ (\texttt{Nadj} or \texttt{N\_adj} in the plot).}
\end{figure}

\subsubsection{Caveat}\label{sec-final-caveat}

The use of this final model for converting
MaxN counts to expected absolute abundance is possibly
handicapped by the lack of data (21 boat trips only) and of spatial variability used to train the
calibration approach.

\section{OPERATIONALIZING OUR CALIBRATION MODEL FOR UNPAIRED CAMERA DEPLOYMENTS}\label{sec-formulae}

Again, a regional survey in the project deploys at most one camera type, sometimes without an echosounder. For example, given a trap camera MaxN from the survey, we wish to couple it with the final model in Section~\ref{sec-final-model} to produce an estimate of the expected GAJ abundance, with uncertainty. A fully rigorous approach may be to refit the final model to an augmented dataset, one that includes the original 21 $(i,j,k)$ combinations of $\{y_{ijkD}, y_{ijkS}, y_{ijkT}, y_{ijkR}, N_{ijk}, r_{ijk}\}$, with an additional 22nd that is $\{\text{\texttt{NA}}, \text{\texttt{NA}}, y_T^\ast, \text{\texttt{NA}}, \text{\texttt{NA}}, r^\ast\}$, where a `$\ast$' denotes the survey observation. However, this fully rigorous approach can be impractical, for two reasons: 1) the indexing of $(i,j)$ for the calibration experiment does not apply to the survey (with $i$ denoting boat type that was specific to the experiment), so that the 22nd boat trip is incompatible with the final model; and 2) even if $\nu_{x,i}$ were set at 0 in the final model (i.e., boat type were assumed to not affect absolute abundance) to restore compatibility, model refitting is highly computationally intensive for use during a field survey.  

A more practical alternative for use in the field is to derive a set of calibration formulae that take on a camera-specific MaxN as input and provide an estimated expected GAJ abundance as output. To this end, we conduct least-squares regression similar to that described in Section~\ref{sec-final-model}, as follows.

For each \(M\) and each \(\ell(=D,S,T,R)\), the least-squares regression line of \(\{\phi_s^{(M)}: s=1,...,21\}\) on \(\{y_{s\ell}:s=1,...,21 \text{ and } y_{s\ell} \text{ not missing}\}\) has intercept 
\(b_{0,\ell}^{(M)}\) and slope \(b_{1,\ell}^{(M)}\). Thus, the corresponding linear formula can take an $\ell$th camera-specific MaxN count as input and map it to the \(M\)th estimated expected absolute abundance of GAJ as output.
Next, fixing \(\ell\), we gather
\((b_{0,\ell}^{(M)}, b_{1,\ell}^{(M)})\) across all values of \(M\) to
form the joint posterior distribution of the \textit{calibration
intercept and slope} for the \(\ell\)th camera type. The
posterior median intercept and posterior median slope can be used to
produce \emph{the} calibration formula for converting an $\ell$th MaxN count to an
absolute abundance estimate.

Figure~\ref{fig-widget} shows screen shots of the interactive R Shiny app \citep{shiny} that we have created to display posterior summaries and
visualizations of the calibration formulae for cases in which a camera is deployed alone (without being accompanied by the echosounder). The app user chooses the camera type to see the posterior median, mean, and variance for each of $b_0$ and $b_1$ from least-squares regression of posterior draws of $\boldsymbol{\phi}$ on this camera type's MaxN data in the calibration experiment; additionally, the posterior correlation between $b_0$ and $b_1$ is shown. The app user can use these posterior statistics to manually convert a regional survey's MaxN count to an estimated expected GAJ abundance and compute an approximate standard error (SE). Alternatively, the widget displays an interactive plot of 5000 least-squares lines (blue) and the ``posterior median'' least-squares line (white); the app user can click on the white line at a particular MaxN value to see the corresponding estimated value of expected GAJ abundance; hovering the mouse over any part of the plot gives the coordinates of the mouse, which facilitates the visual evaluation of the approximate SE.

\begin{figure}
\centering{
\includegraphics[scale=.15]{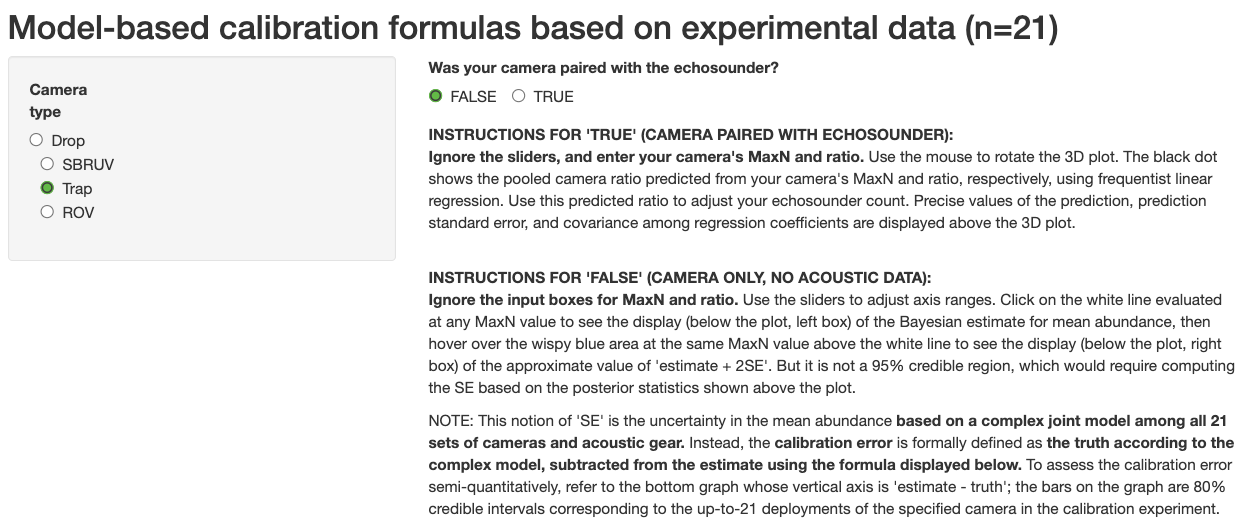} $\quad$
\includegraphics[scale=.15]{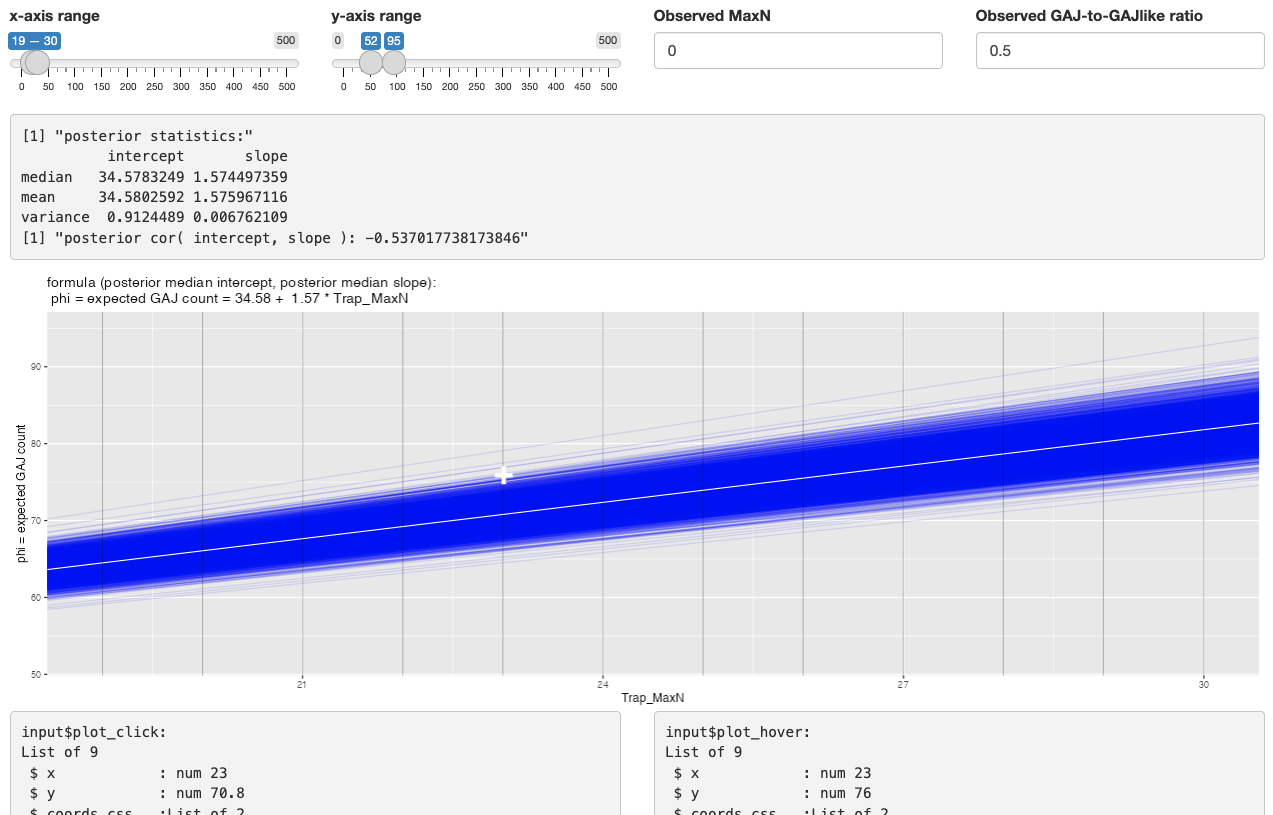} $\quad$
\includegraphics[scale=.12]{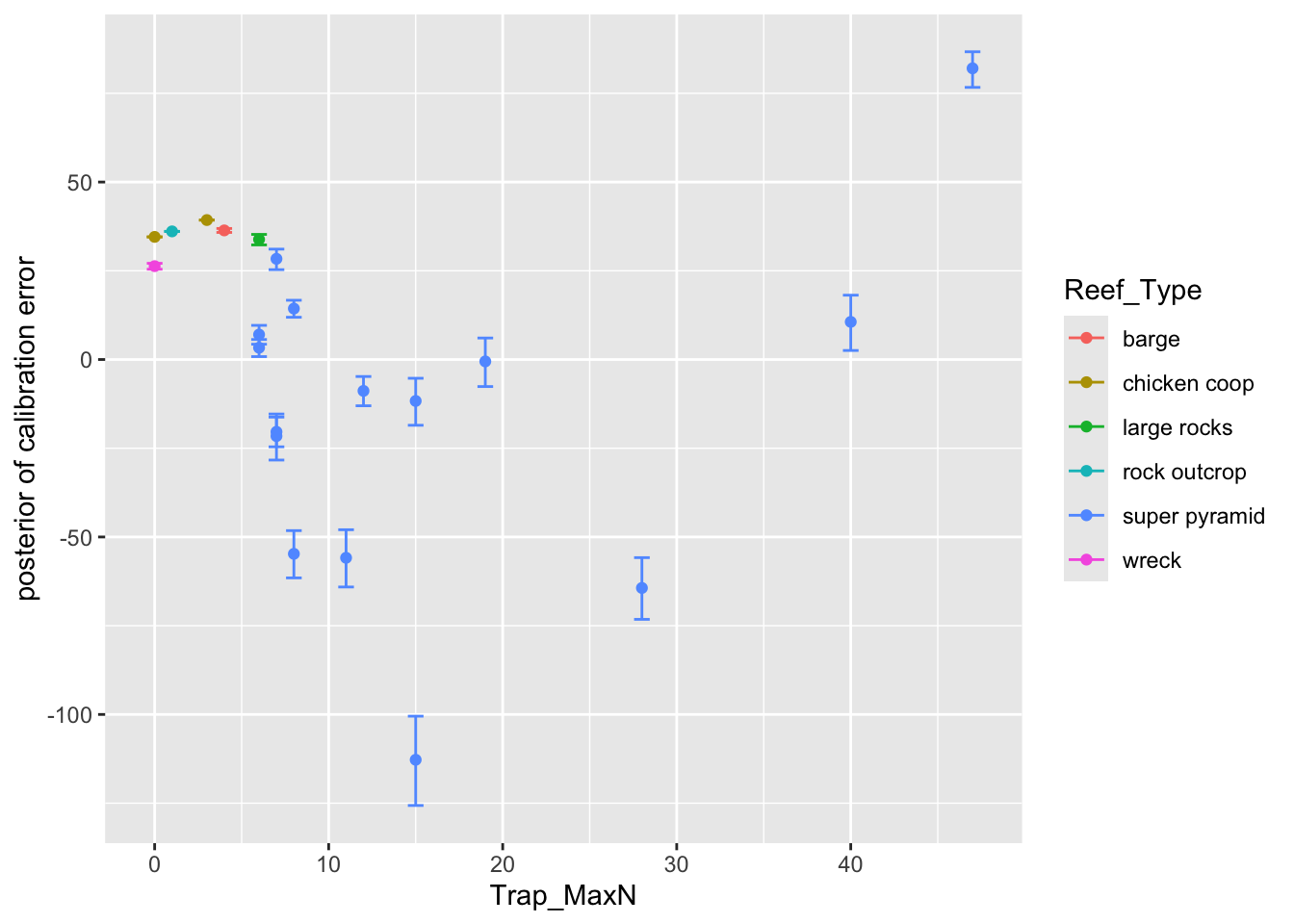}
}
\caption{\label{fig-widget}\textbf{Calibration widget instructions and
interface for unpaired MaxN.} Top-left panel: Instructions for using the widget interface in
the case of either an unpaired MaxN (single camera type, unaccompanied by echosounder) or a MaxN paired with an acoustic count. Top-right panel: Interactive widget interface with calibration formula and associated output for an unpaired MaxN after
changes have been made to the default input values; the white ``+''
denotes the location of the hovering mouse. Camera-specific posterior
summaries for the corresponding Bayesian-model based calibration formula
are included in the display. They include the posterior mean, median,
and variance for each of the calibration intercept and slope. Also
included is the posterior correlation between the calibration intercept
and slope. 
Bottom panel: Widget output for the \textit{calibration error} evaluated at each observation in the calibration experiment.}
\end{figure}

Note that the notion of this SE corresponds to the uncertainty in the $\boldsymbol{\phi}$ vector associated with the final model in Section~\ref{sec-final-model}. Specifically, each posterior draw of $\boldsymbol{\phi}$ from the final model yields a different regression line (in blue in Figure~\ref{fig-widget}). Thus, this SE does not reflect the goodness-of-fit of the calibration formula (white line in Figure~\ref{fig-widget}). For example, the posterior median $R^2$ value for the $\phi$-on-$y_T$ least-squares regression is 0.16, reflecting a poor goodness-of-fit, and hence, a possibly large \textit{calibration error}. To quantify this calibration error with more rigor, the widget displays the posterior median and 80\% credible interval for the quantity $\hat{\phi}_{ijk}-\phi_{ijk}$ for each $(i,j,k)$, where $\hat{\phi}_{ijk}$ is the value according to the calibration formula (white line in Figure~\ref{fig-widget}). For a given survey value of, say, $y_T^\ast$, the user can match the reef type associated with $y_T^\ast$ to those reef types in the calibration error display, and gauge the calibration error due to employing the $y_T$-to-$\phi$ calibration formula on $y_T^\ast$.

\section{DATA RICH SCENARIOS --- A SIMULATION STUDY}\label{sec-sim}

In a larger-scale calibration experiment with more data / fewer missing data / a more balanced experimental design, our most comprehensive calibration model may be desirable. Here, we conduct a simulation study in which the model in Section~\ref{sec-complex-model} is used to generate data, while fixing population-level parameters at values that are inspired by the results from Section~\ref{sec-final-model}. The computer implementation of assigning parameter values and simulating data can be found in the Supplementary Materials.\footnote{Computer code associated with this manuscript will be submitted upon its acceptance for publication.} An overview of the procedure is as follows.  

First, we directly adopt from our 21 boat trips the data on vessel, reef size, pooled ratio $r$, MaxN's, and acoustic and mark-recapture counts. We also extract $\widehat{\log\phi}$ (= posterior median values for $\log\phi$) from Section~\ref{sec-final-model}. Values assigned for $\beta_0, \nu_{x,i}, \gamma_{x,j},$ and $\sigma_{\phi,j}$ in Level 3 of the simulation are inspired by a least-squares regression of $\widehat{\log\phi}$ on vessel and reef size. Next, $\{\xi_h+\zeta_{hijk}\}$ in Level 2.2 is approximately $\{\log(r_{ijk} N_{ijk}+1)-\widehat{\log\phi}_{ijk}\}$ for $h=1$ and $\{\log(N_{ijk}^{(mr)}+1)-\widehat{\log\phi}_{ijk}\}$ for $h=2$, which can inspire the assigned values for $\sigma_{x,h}>0$ for all $h$, and for $\xi_h$ (subject to a sum-to-zero constraint). Similarly, for $\ell\ne R$ in Level 2.1, a least-squares regression of $\log y (\approx \log\mu)$ on $\widetilde{\log\phi}$ (computed based on $\widehat{\log\phi}$), vessel, and reef size can inspire the assigned values of $\beta_{y,0,\ell}$ and $\beta_{1,j\ell}$, which are in turn used to approximate $\{\nu_{i\ell}+\gamma_{j\ell}+\varepsilon_{ijk\ell}\}$ to further inspire $\sigma_y$, as well as $\nu_{i\ell}$ and $\gamma_{j\ell}$ (both subject to sum-to-zero constraints). For $\ell=R$ in Level 2.1, we again take $\log y_{ijk\ell}\approx\log\mu_{ijk\ell}$, and regress $\log y_{\ell=R}$ on $\widetilde{\log\phi}, \widetilde{\log y}_{\ell=D}, \widetilde{\log y}_{\ell=S}, \widetilde{\log y}_{\ell=T},$ and reef size to inspire the assigned values of $\beta_{1,jR},\beta_{1,S},$ and $\beta_{1,T}$. However, complications result from camera types $D$ and $R$ never being paired at small reef structures, so that a regression of $\{\log y_{\ell=R}-\beta_{1,1R}\widetilde{\log\phi}-\beta_{1,S}\widetilde{\log y}_{\ell=S}-\beta_{1,T}\widetilde{\log y}_{\ell=T}\}$ on $\widetilde{\log y}_{\ell=D}$ is separately fitted for large reefs only to inspire the assigned value of $\beta_{1,D}$. All these assigned slope values can be substituted back into Level 2.1 in the model to inspire the assigned values for $\beta_{y,0,R}, \gamma_{y,jR},$ and $\sigma_{yR}$. If at any step a least-squares slope coefficient is negative, it is first multiplied by $-1$ before being used to inspire other model parameters.

The final step before simulating data is to jitter the pooled camera ratio $r$ on the log scale by a standard deviation of approximately 0.27. These and all assigned parameters are then used to simulate 126 boat trips on MaxN ($y$), acoustic count ($N$), and mark-recapture estimates ($N^{(mr)}$) according to Levels 1--3 in the model. The predictor variables in each set of 126 ($=21\times 6$) boat trips is made up of replicating 6 times the original 21 observations on vessel, reef size, and camera ratio, with the exception that the camera ratios are jittered instead of cloned.

We simulate 50 datasets of 126 boat trips in this manner, then fit the comprehensive calibration model to each dataset, again using NIMBLE and visual assessment of MCMC convergence. We extract nominal 90\% credible intervals for each population-level parameter in the model, and record the capture rate out of 50. We additionally monitor the capture rate for the derived parameters $\mathcal{M}_{ij}^{\log}=E(\log\phi_{ijk})=\beta_0+\nu_{x,i}+\gamma_{x,j}$ and $\mathcal{M}_{ij}=E(\phi_{ijk})=\exp\{\mathcal{M}_{ij}^{\log} + \sigma_{\phi,j}^2/2\}$.

\subsection{RESULTS}  

Table~\ref{tab-sim} presents the capture rate of nominal 90\% credible intervals for all population-level parameters, as well as $\mathcal{M}_{ij}^{\log}$ and $\mathcal{M}_{ij}$. 

\begin{table}[htbp]
\caption{\textbf{Simulation results.} Capture rate (out of 50 sets) of nominal 90\% credible intervals for all population-level parameters and associated derived parameters, $\mathcal{M}_{ij}^{\log}$ and $\mathcal{M}_{ij}$, for our comprehensive calibration model (Section~\ref{sec-complex-model}).\label{tab-sim}}
\begin{center}
\begin{tabular}{llc|llc|llc}
  \hline
Level & Parameter & Rate & Level & Parameter & Rate & Level & Parameter & Rate \\
\hline
2.1 & $\beta_{y,0,D}$ & 0.04 & 2.1 (R) & $\beta_{y,0,R}$ & 0.46 & 3 & $\beta_0$ & 0.98\\
(D/S/T) & $\beta_{y,0,S}$ & 0.04 & & $\beta_{1,1R}$ & 0.00  &  & $\nu_{x,1}$ & 0.87\\
& $\beta_{y,0,T}$ & 0.24 & & $\beta_{1,2R}$ & 0.00  & & $\gamma_{x,1}$ & 0.93\\
& $\beta_{1,1D}$ & 0.00  & & $\beta_{1,D}$ & 0.70 & & $\sigma_{\phi,1}$ & 0.83\\
& $\beta_{1,1S}$ & 0.00 & & $\beta_{1,S}$ & 0.83 & & $\sigma_{\phi,2}$ & 0.87\\
& $\beta_{1,1T}$ & 0.02 & & $\beta_{1,T}$ & 0.00 & & $\mathcal{M}_{11}$ & 0.87\\
& $\beta_{1,2D}$ & 0.37 & & $\gamma_{y,1R}$ & 0.96 & & $\mathcal{M}_{21}$ & 0.91\\
& $\beta_{1,2S}$ & 0.70 & & $\sigma_{yR}$ & 0.13 & & $\mathcal{M}_{12}$ & 0.89\\
& $\beta_{1,2T}$ & 0.98 & &                         & & & $\mathcal{M}_{22}$ & 0.89\\
& $\nu_{y,1D}$ & 0.67 & 2.2 & $\xi_1$ & 0.93 & & $\mathcal{M}_{11}^{\log}$ & 0.96\\
& $\nu_{y,1S}$ & 0.57 & & $\sigma_{x,1}$ & 0.89 & & $\mathcal{M}_{21}^{\log}$ & 0.85\\
& $\nu_{y,1T}$ & 0.57 & & $\sigma_{x,2}$ & 0.85 & & $\mathcal{M}_{12}^{\log}$ & 0.96\\
& $\gamma_{y,1D}$ & 0.89 & &                    & & & $\mathcal{M}_{22}^{\log}$ & 0.96\\
& $\gamma_{y,1S}$ & 0.09 & &                    & & & & \\
& $\gamma_{y,1T}$ & 0.52 & &                    & & & & \\
& $\rho$ & 0.50 &          &                    & & & & \\
& $\sigma_y$ & 0.46 & & & & & Combined & 0.59 \\ \hline
\end{tabular}  
\end{center}
\end{table}

While the capture rate for each parameter in Levels 2.2 and 3 is near or exceeds the nominal rate of 90\%, that for many parameters in Level 2.1 is as poor as 0\%, although 6 of the Level 2.1 parameters exhibit capture rates between 0.70 and 0.98. The overall capture rate (for all parameters combined, including the derived parameters) is just shy of 0.60. These results may suggest that the upper levels in the model hierarchy are poorly identified. Notwithstanding, the key purpose of the calibration model is to reconcile gear-specific MaxN's, acoustic data, and mark-recapture estimates for the unified inference of $\phi$, the expected count of the target species. In that sense, the model performs very well, whereby both $\mathcal{M}_{ij}^{\log}=E(\log\phi)$ and $\mathcal{M}_{ij}=E(\phi)$ are successfully recovered at a rate that is near or above the nominal capture rate.

\section{ADJUSTING ACOUSTIC DATA PAIRED WITH CAMERA}\label{sec-adjust}

For most regional surveys, a single camera type is deployed alongside
acoustic gear. In these cases, inference of $\phi$ should account for the joint presence of $N$ and $y_\ell$ in the survey. Thus, an immediate concern becomes the derivation of an appropriate adjustment factor $r$ to correct $N$. However, each camera type may have its unique limitations, so that an adjustment factor based on the single camera type deployed in the survey may be biased. A more robust adjustment factor is needed in this case.

As seen in Figure~\ref{fig-camratios}, the pooled ratio \(r_s\) from our
calibration experiment aligns all four camera types against each other,
so that \(r_s\) as an adjustment factor for acoustic counts should be
more robust than any individual camera-specific ratio. As such, for a
camera-specific ratio and MaxN observed in a field survey that also deploys the echosounder, we use the ratio and MaxN to predict a pooled \(r_s\); our prediction
is based on a least-squares regression fit of \(r_s\) on the ratio and MaxN from
the calibration experiment.

Figure~\ref{fig-camratios} suggests that \(r_s\) depends rather linearly
on each camera-specific ratio. Figure~\ref{fig-linreg} further suggests
a linear relationship between the camera-specific ratio and the square of
log(MaxN + 1). Thus, the mean of \(r_s\) appears to be a quadratic
function of log(MaxN + 1); due to collinearity between the square term
and the camera ratio, the quadratic function can be re-expressed as a
linear combination of log(MaxN + 1) and the camera ratio.

\begin{figure}
\centering{
\includegraphics[scale=.14]{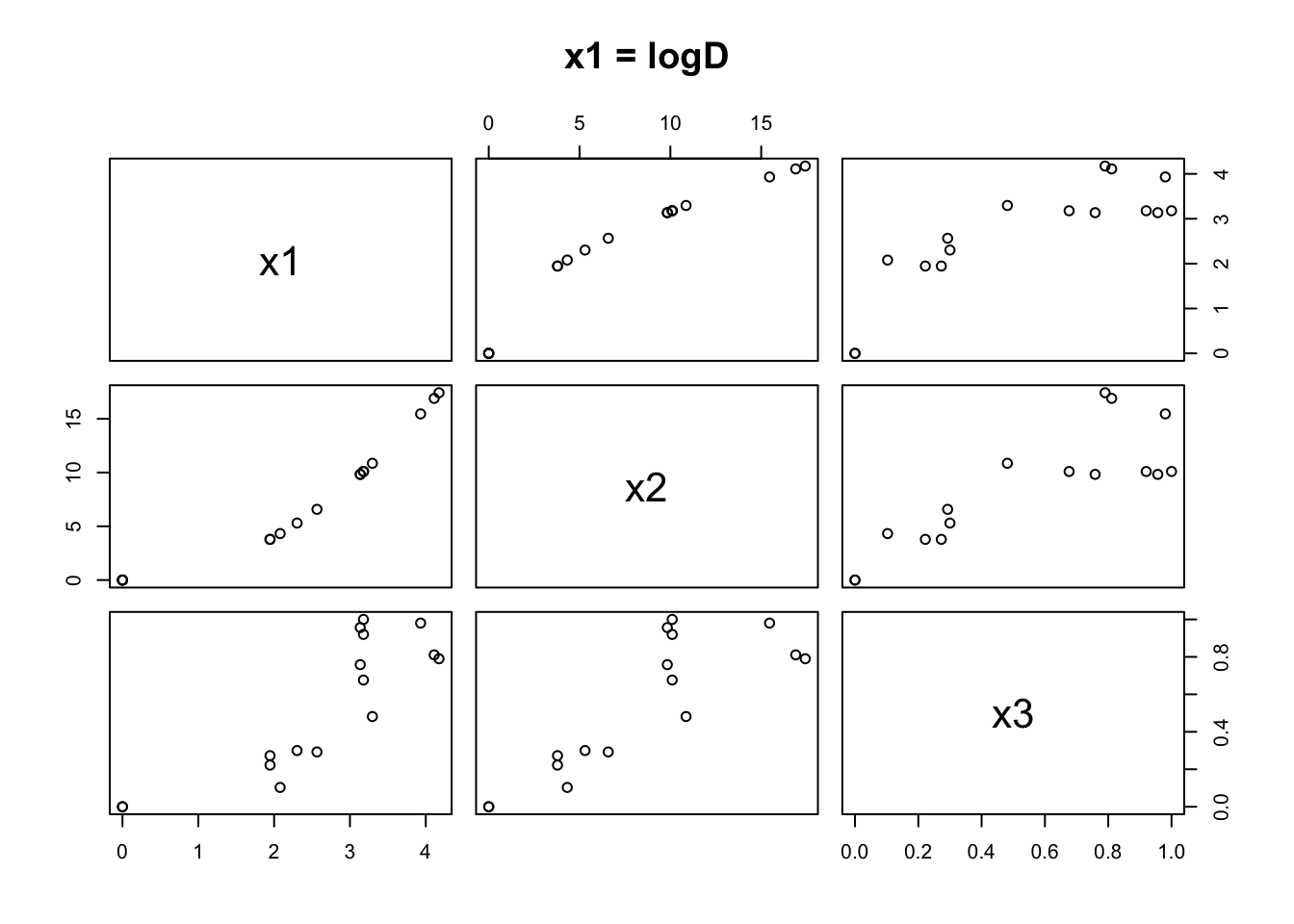} $\quad$
\includegraphics[scale=.14]{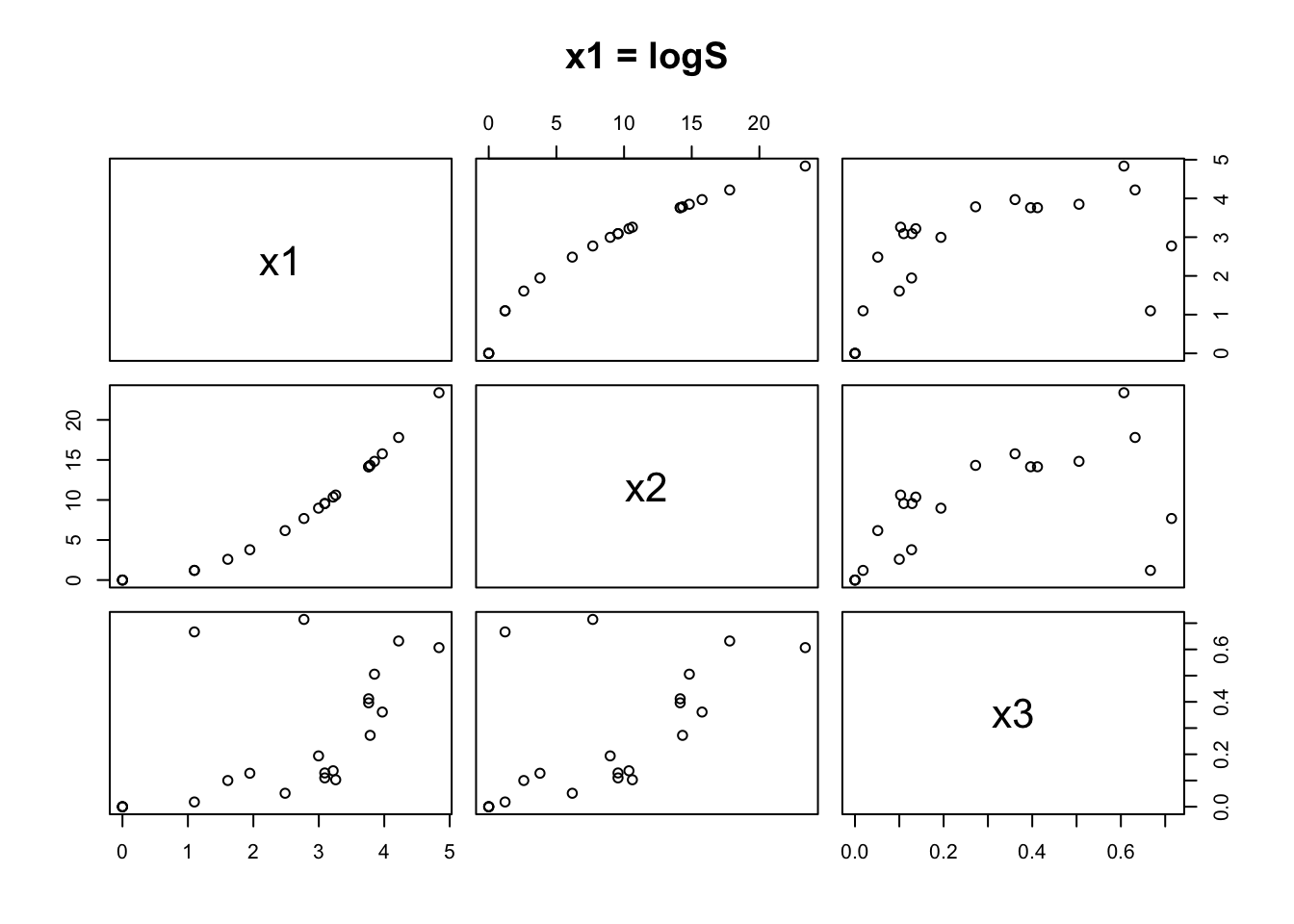}

\includegraphics[scale=.14]{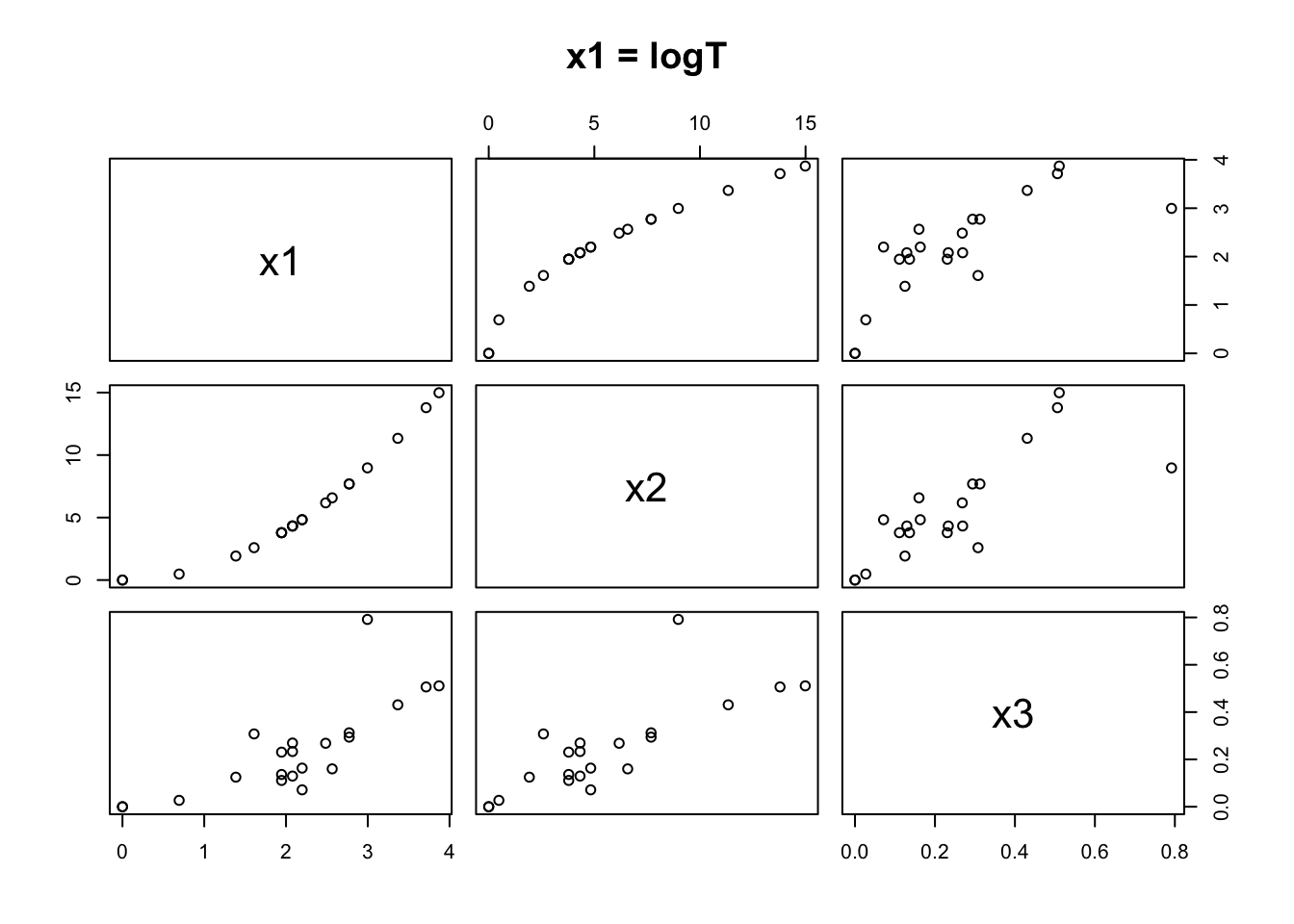} $\quad$
\includegraphics[scale=.14]{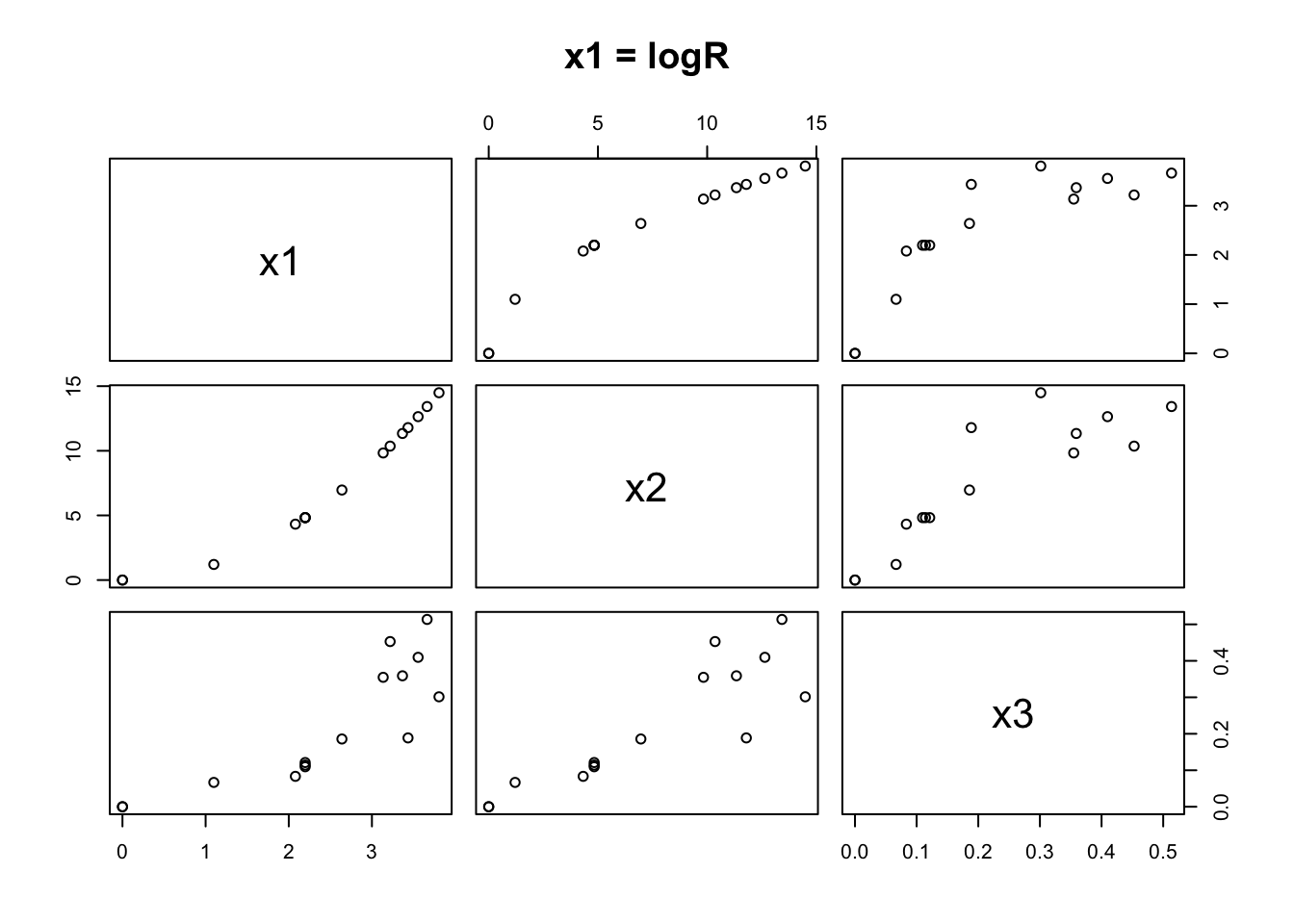}
}
\caption{\label{fig-linreg}\textbf{Scatter-plot matrices} among
\texttt{x1} = log(MaxN + 1), \texttt{x2} = (\texttt{x1})\(^2\), and
\texttt{x3} = camera ratio for drop camera (top left panel), SBRUV
camera (top right panel), trap camera (bottom left panel), and ROV
camera (bottom right panel).}
\end{figure}

Finally, homogeneity in Figures~\ref{fig-camratios} and
\ref{fig-linreg}, together with the trends described above,
suggest that a least-squares linear regression of \(r_s\) on log(MaxN + 1)
and camera ratio can be used to predict \(r_s\) in this case. 

The resulting least-squares fits are operationalized into a
prediction tool as part of our R Shiny widget
(Figure~\ref{fig-widget2}). The app user chooses the camera type to display the least-squares regression coefficients and their covariance matrix based on the calibration experiment. The app user can use these statistics to manually compute an $\hat{r}$ (i.e., the predicted pooled-camera ratio) and prediction standard error given any MaxN count and associated camera-specific ratio of GAJ to GAJ+. Alternatively, the app user can enter an observed value for each of MaxN and the camera-specific GAJ:GAJ+ ratio. The widget then displays $\hat{r}$ and the associated prediction standard error.

\begin{figure}
\centering{
\includegraphics[scale=.24]{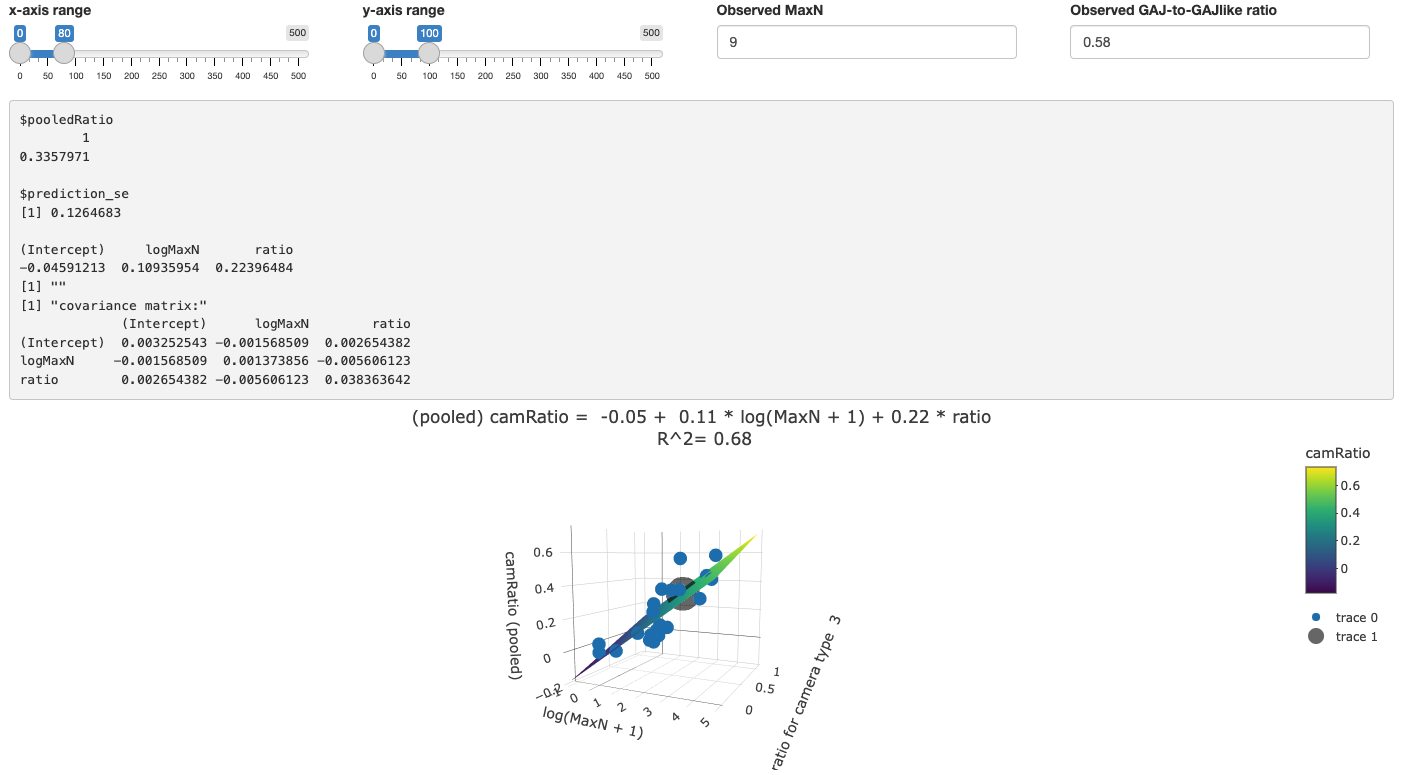}
}
\caption{\label{fig-widget2}\textbf{Calibration widget interface for camera paired with echosounder.}
Interface with output after changes have been made to the default
input values. An interactive 3D plot displays the least-squares
regression of pooled camera ratio on the camera-specific MaxN (log
scale) and ratio, respectively. The predicted \(\hat{r}\) for the pooled
camera ratio (large black dot in the 3D plot) and associated prediction
standard error are provided based on the user's camera-specific MaxN and
ratio. Also included are the regression coefficients, \(R^2\), and
variance-covariance matrix among the regression coefficients.}
\end{figure}

\section{CONCLUSION AND FUTURE WORK}\label{sec-tbd}

To faciliate the inference for absolute fish species abundance at a continental scale, in this paper, we have presented a methodology for reconciling gear-specific data on MaxN, a relative abundance index, that are collected in regional surveys. The methodology requires data from a calibration study in which multiple region-specific camera types are simultaneously deployed alongside an echosounder, the latter of which records the combined abundance of GAJ+ species on the absolute scale. Data from the calibration study allow us to construct 1) a Bayesian hierarchical framework that jointly models all MaxN and echosounder data to produce camera-specific calibration formulae that can convert a MaxN to an estimate of mean GAJ absolute abundance; 2) a least-squares regression framework that jointly models all MaxN data and camera ratios of GAJ:GAJ+ to produce camera-specific formulae that can take a MaxN and camera ratio as input, giving as output an estimate of a non-camera-specific ratio to be used to deflate an echosounder count from GAJ+ abundance to GAJ abundance. Both types of conversions are accompanied by uncertainty quantification, and are operationalized via a single R Shiny widget. Conversion 1 is to be used in a regional survey that deploys a single camera type that is unpaired with an echosounder, while Conversion 2 is used when a single camera type is paired with an echosounder. Through simulations and cross-validation against mark-recapture data, we have shown our methodological framework for Conversion 1 to be a promising approach that may be adapted to other continental-scale species abundance studies which deploy regional detection gears that yield data on various relative abundance scales, thus providing a more holistic picture on the actual abundance of the species of interest. We believe that our methodology can help to advance the research and policy-making communities' efforts towards meeting sustainable development goals. 

We must note some limitations in the use of our methodology for the GAJ project.

For the Bayesian modeling framework (Conversion 1), we have used data from the 21 boat trips in a very small-scale calibration experiment to develop the calibration model that allows unpaired, camera-specific MaxN (relative abundance) counts from a regional survey to be converted to estimates of expected absolute abundance, accompanied by uncertainty estimates. As discussed in Section~\ref{sec-final-caveat}, these calibration formulae for the GAJ project may lack robustness across the entire region of
study, because they are constructed using a small dataset with small spatial coverage. Moreover, the current model ignores the notion of a
``denominator'' in the definition of abundance density. In particular,
none of the MaxN's or acoustic counts in this model has been adjusted
for the spatial coverage associated with each gear, as this
information is not yet available. Because the true mean abundance \(\phi\) is
modeled as the expectation of the ratio-adjusted acoustic count, it shares the
same spatial denominator as the acoustic count \(N\), which may differ
across habitat structures. Thus, the final model in
Section~\ref{sec-final-model} may need to be updated by incorporating areal coverage information as they become
available.

In the case of joint deployment of a single camera type and the acoustic echosounder in a regional survey (Conversion 2), we have developed an adjustment approach for \(N\) that is anticipated to be more robust
than a GAJ:GAJ+ ratio derived from the single deployed camera type. Our
least-squares regression serves this purpose by producing a
predicted pooled ratio, but it may suffer from the same lack of spatial robustness as Conversion 1. Moreover, it remains unclear if the resulting adjusted value $(=\hat{r}N)$
should be regarded as the estimated absolute abundance without
further modeling. For example, it may be desirable to combine the 21
experimental data points with the field survey's MaxN and
regression-adjusted \(N\) to produce an updated Bayesian hierarchical
model for the true mean abundance \(\phi\). The resulting updated model
could be a more realistic model for the unpaired case. However, in
such an updated model, all field survey trips will have missing data for
the undeployed camera types. Without actually fitting the updated
Bayesian model, it is unclear if the enormous amount of missing data may
hinder the inference for \(\phi\).

Finally, C-BASS calibration will be handled independently of the
approach developed in this paper.

\begin{appendix}
\SupplementaryMaterials
\section{Timeline of calibration modeling}\label{sec-appdx-timeline}

\subsection{Cross calibrating camera counts
only}\label{models-v1-cross-calibrating-camera-counts-only}

Model variants based on the following were fitted for exploratory purposes:  

\begin{align*}
\text{Level 1:} && y_{ijk\ell} &\sim \text{Pois}(\mu_{ijk\ell}) \\\\
\text{Level 2:} && \text{for } \ell\ne R: \quad \log\mu_{ijk\ell} &= \beta_{y,0,\ell} + \nu_{y,i\ell} + \gamma_{y,j\ell} + \beta_{1,j\ell}\log\phi_{ijk} + \varepsilon_{ijk\ell} \\
&& \left[ \begin{matrix}
\varepsilon_{ijkD} \\
\varepsilon_{ijkS} \\
\varepsilon_{ijkT} 
\end{matrix} \right] &\sim \text{MVN}(\mathbf{0}, \sigma_y^2\mathbb{P}) \\ 
&& \mathbb{P}_{\ell_1,\ell_2} &= \mathbb{P}_{\ell_2,\ell_1} = \rho_{\ell_1,\ell_2} \text{ for all } \ell_1\ne\ell_2 \text{ and }\ell_1,\ell_2 = D,S,T \\
&& \text{for } \ell=R: \quad \log\mu_{ijkR} &= \beta_{y,0,\ell} + \gamma_{y,jR} + \beta_{1,jR}\log\phi_{ijk} + \\
&& &\quad\quad \beta_{1,D}\log\mu_{ijkD} + \beta_{1,S}\log\mu_{ijkS} + \\
&& &\quad\quad\quad \beta_{1,T}\log\mu_{ijkT} + \varepsilon_{ijkR} \\
&& \varepsilon_{ijkR} &\sim \mathcal{N}(0, \sigma_{yR}^2) \\
&& \text{for all } \ell: \quad\quad 0 = \sum_i \nu_{y,i\ell} &= \sum_j \gamma_{y,j\ell}\quad \text{for identifiability} \\
\text{Level 3:} && \log\phi_{ijk} &= \beta_0 + \nu_{x,i} + \gamma_{x,j} + \delta_{ijk} \\
&& \delta_{ijk} &\sim \mathcal{N}(0,\sigma_{\phi,j}^2) \\
&& 0 = \sum_i \nu_{x,i} &= \sum_j \gamma_{x,j} \quad \text{for identifiability}
\end{align*}
where $\beta_{y,0,\ell}\equiv 0$ to reduce confounding, and \(\rho_{\ell_1,\ell_2}\) may not be constant across
\(\ell_1,\ell_2 = D,S,T\).

\subsubsection*{NOTES}

\begin{itemize}
\item
  The purpose of these exploratory models was not true calibration.
\item
  We decided to force \(\rho_{\ell_1,\ell_2}\) to be constant for all
  future versions.
\item
    Further model refinement for these exploratory models was not pursued.
\end{itemize}

\subsection{Mark-recapture counts (GAJ only) and ``raw''
acoustic counts (all GAJ+ species, unadjusted)}\label{models-v2v3-mark-recpature-counts-gaj-only-and-raw-acoustic-counts-all-gaj-like-fish}

\begin{enumerate}
\item \label{MR1} Model variants based on the following were fitted to incorporate mark-recapture (MR) data as true absolute abundance observations, i.e., we calibrated MaxN's against expected MR count values. However, camera counts were observed on both \texttt{Reef\_Size} values
    (\texttt{L}, \texttt{S}), yet MR counts were observed at
    \texttt{Reef\_Size=L} only. Thus, MR counts were
    further calibrated against a latent ``true expected abundance'' that could depend on \texttt{Reef\_Size}.
      \begin{align*}
      \text{Level 1.1:} && y_{ijk\ell} &\sim \text{Pois}(\mu_{ijk\ell}) \\
      \text{Level 2.1:} && N_{i1k}^{(mr)} &\sim \text{Pois}(\phi_{i1k}) \quad \text{for all } i,k \quad (j=1 \text{ only}) \\
      \text{Level 2.2:} && \text{for } \ell\ne R: \quad \log\mu_{ijk\ell} &= \nu_{y,i\ell} + \gamma_{y,j\ell} + \beta_{1,j\ell}\log\phi_{ijk} + \varepsilon_{ijk\ell} \\
      && \left[ \begin{matrix}
      \varepsilon_{ijkD} \\
      \varepsilon_{ijkS} \\
      \varepsilon_{ijkT} 
      \end{matrix} \right] &\sim \text{MVN}(\mathbf{0}, \sigma_y^2\mathbb{P}) \\ 
      && \mathbb{P}_{\ell_1,\ell_2} &= \mathbb{P}_{\ell_2,\ell_1} = \rho \quad \text{ for all } \ell_1\ne\ell_2 \text{ and }\ell_1,\ell_2 = D,S,T \\
      && \text{for } \ell=R: \quad \log\mu_{ijkR} &= \gamma_{y,jR} + \beta_{1,jR}\log\phi_{ijk} + \\
      && &\quad\quad \beta_{1,D}\log\mu_{ijkD} + \beta_{1,S}\log\mu_{ijkS} + \\
      && &\quad\quad\quad \beta_{1,T}\log\mu_{ijkT} + \varepsilon_{ijkR} \\
      && \varepsilon_{ijkR} &\sim \mathcal{N}(0, \sigma_{yR}^2) \\
      && \text{for all } \ell: \quad\quad 0 = \sum_i \nu_{y,i\ell} &= \sum_j \gamma_{y,j\ell}\quad \text{for identifiability} \\
      \text{Level 3:} && \log\phi_{ijk} &= \beta_0 + \nu_{x,i} + \gamma_{x,j} + \delta_{ijk} \\
      && \delta_{ijk} &\sim \mathcal{N}(0,\sigma_{\phi,j}^2) \\
      && 0 = \sum_i \nu_{x,i} &= \sum_j \gamma_{x,j} \quad \text{for identifiability}
      \end{align*}

\textbf{RESULTS:} Posterior median of \(\log\phi\) and \(\log (N+1)\) were negatively
    correlated at super pyramids (Figure~\ref{fig-MR1}). We also decided to use quadvariate normal in Level 2 in subsequent models to reduce model complexity.
    \begin{figure}[htpb]
\centering{
\includegraphics[width=3.125in]{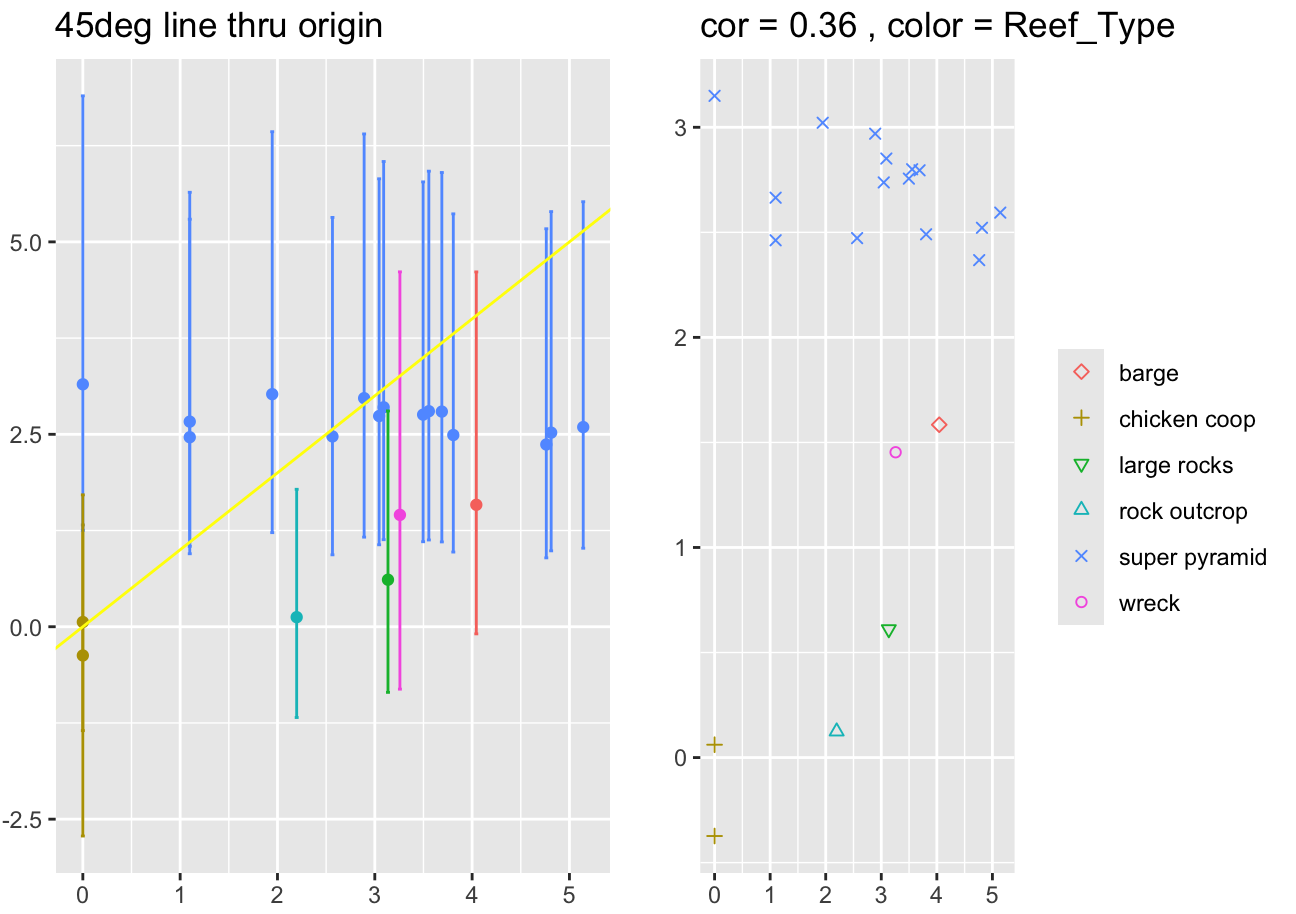}
\caption{Posterior median for $\log\phi$ vs.~ $\log(N+1)$, with 80\% credible intervals (left) and without (right), from one of the best fitting models under Section~\ref{models-v2v3-mark-recpature-counts-gaj-only-and-raw-acoustic-counts-all-gaj-like-fish}, Item \#\ref{MR1}.\label{fig-MR1}}
}
  \end{figure}
\item \label{MR2} We repeated models under \#\ref{MR1} but replaced Level 2.1 with acoustic counts and Level 2.2 with quadvariate normal, as follows:
  \begin{align*}
  \text{Level 2.1:} && N_{ijk} &\sim \text{Pois}(\phi_{ijk}) \quad \text{for all } i,j,k \\
  \text{Level 2.2:} && \left[ \begin{matrix}
      \varepsilon_{ijkD} \\
      \varepsilon_{ijkS} \\
      \varepsilon_{ijkT} \\
      \varepsilon_{ijkR} 
      \end{matrix} \right] &\sim \text{MVN}(\mathbf{0}, \sigma_y^2\mathbb{P}) \\ 
      && \mathbb{P}_{\ell_1,\ell_2} &= \mathbb{P}_{\ell_2,\ell_1} = \rho \quad \text{for all } \ell_1\ne\ell_2 
  \end{align*}
  i.e., we used acoustic counts as data on true absolute abundance, thus calibrating MaxN's against the acoustic expected value.
  
\textbf{RESULTS:} On the log scale, the estimated ``true expectation'' (\(\log\phi\))
    aligned almost perfectly with logged MR counts but
    poorly with logged acoustic counts (Figure~\ref{fig-MR2}). 
    \begin{figure}[!htpb]
\centering{
    \includegraphics[width=2.125in]{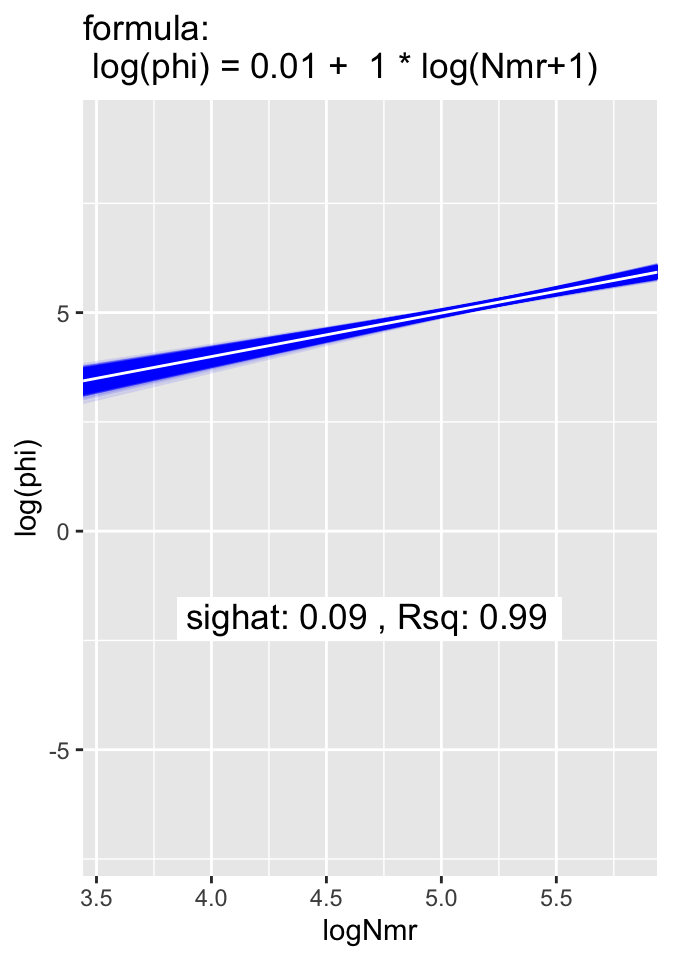} $\quad$ 
    \includegraphics[width=2.125in]{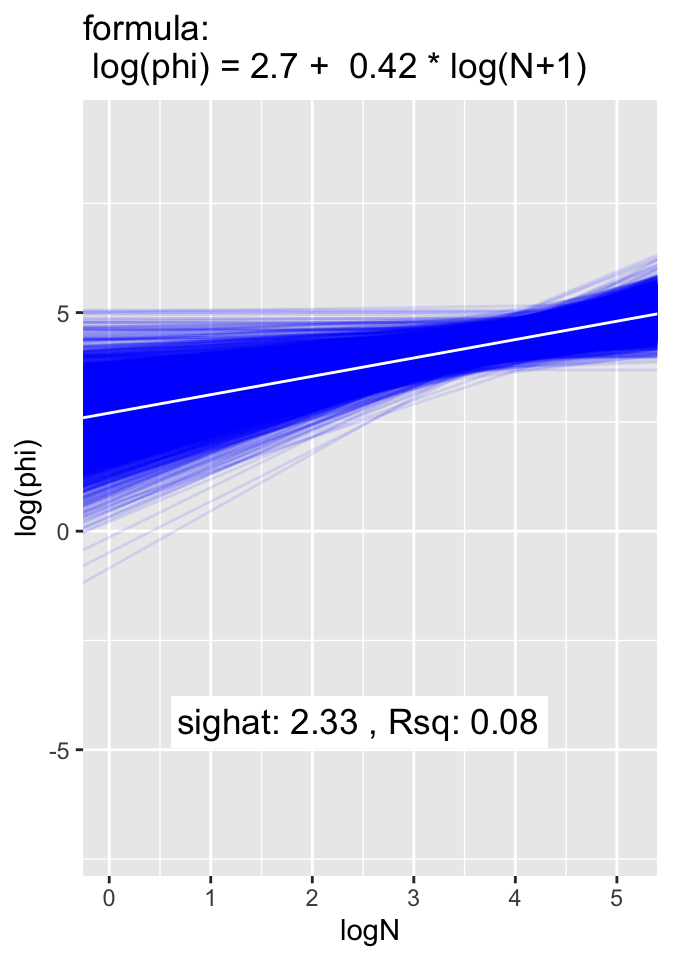}
}
\caption{Posterior draws of $\log\phi$ regressed on $\log(N^{(mr)}+1)$ (left) and $\log(N+1)$ (right), from one of the best fitting models under Section~\ref{models-v2v3-mark-recpature-counts-gaj-only-and-raw-acoustic-counts-all-gaj-like-fish}, Item \#\ref{MR2}; see Figure~\ref{fig-phi-vs-N} for procedural details.\label{fig-MR2}}
  \end{figure}
\vspace{5mm}

\item \label{MR3} We repeated models under \#\ref{MR2} but replaced acoustic counts with updated data (acoustic signals transcribed with a refined algorithm).

\textbf{RESULTS:}
On the log scale, the expected acoustic
      count \(\log\phi\) did not correlate with logged MR counts
      whatsoever, nor did it align well even with \(\log (N+1)\) itself
      (i.e., showed a non-0 intercept and non-unit slope in Figure~\ref{fig-MR3log}). The misalignment may be worse on the count scale (Figure~\ref{fig-MR3}, top row), especially for D- and R-calibrations (Figure~\ref{fig-MR3}, bottom row). 
  \begin{figure}[!htpb]
\centering{
      \includegraphics[width=2.125in]{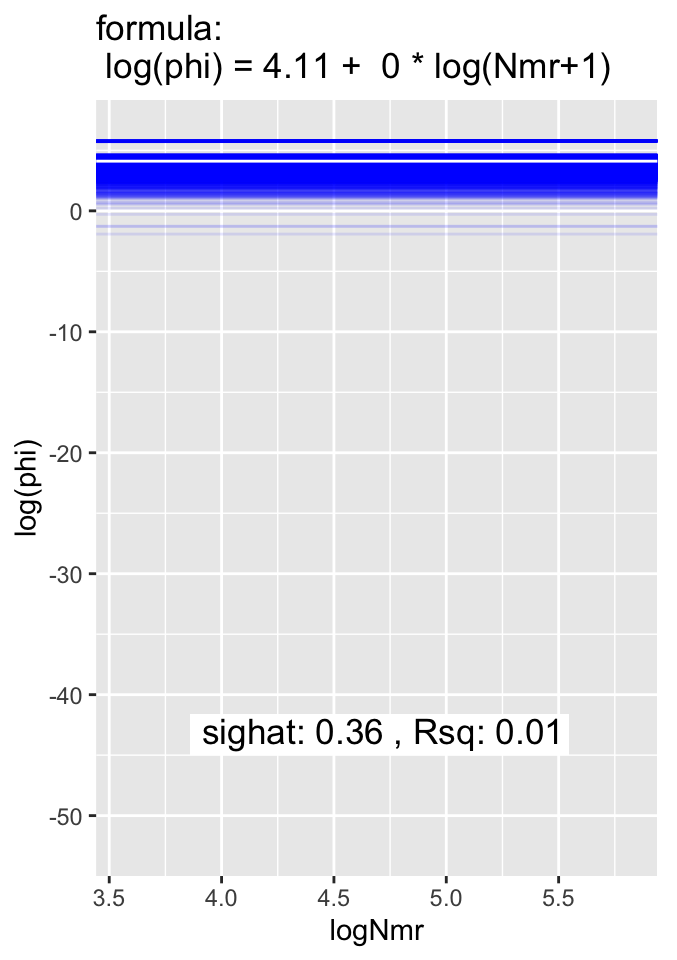} $\quad$
      \includegraphics[width=2.125in]{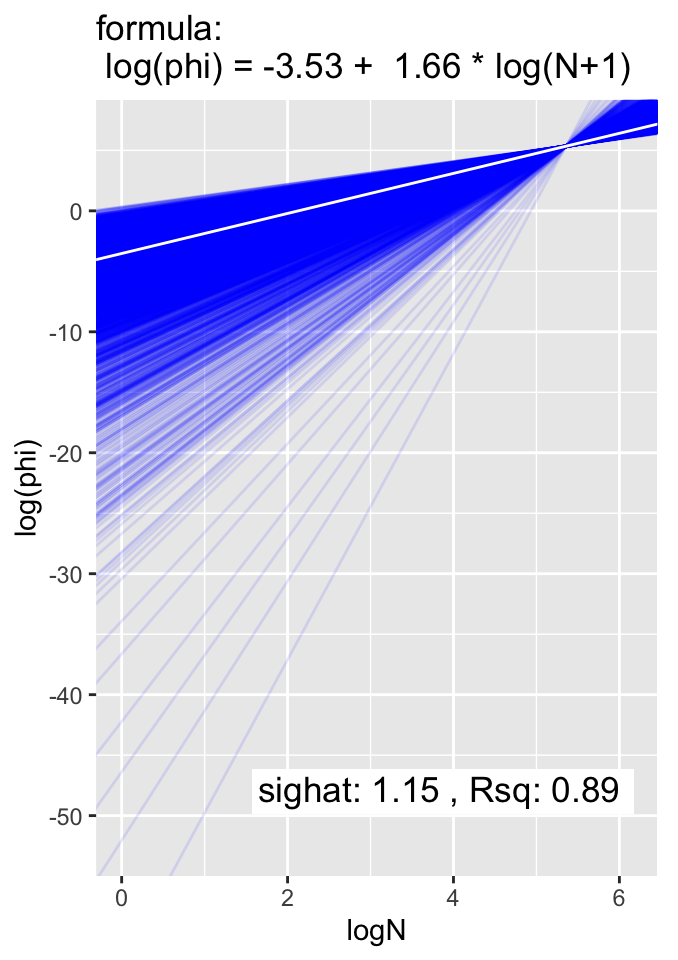}
}
\caption{Same as Figure~\ref{fig-MR2} but from models under Section~\ref{models-v2v3-mark-recpature-counts-gaj-only-and-raw-acoustic-counts-all-gaj-like-fish}, Item \#\ref{MR3}.\label{fig-MR3log}}
  \end{figure}
  \begin{figure}[htbp]
\centering{
      \includegraphics[width=2.125in]{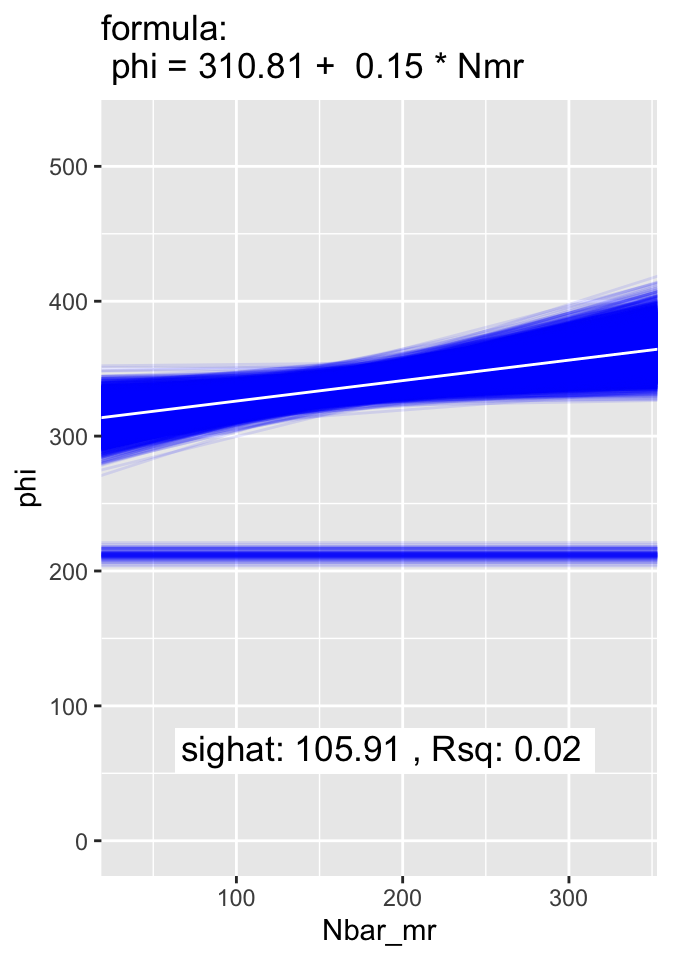} $\quad$
      \includegraphics[width=2.125in]{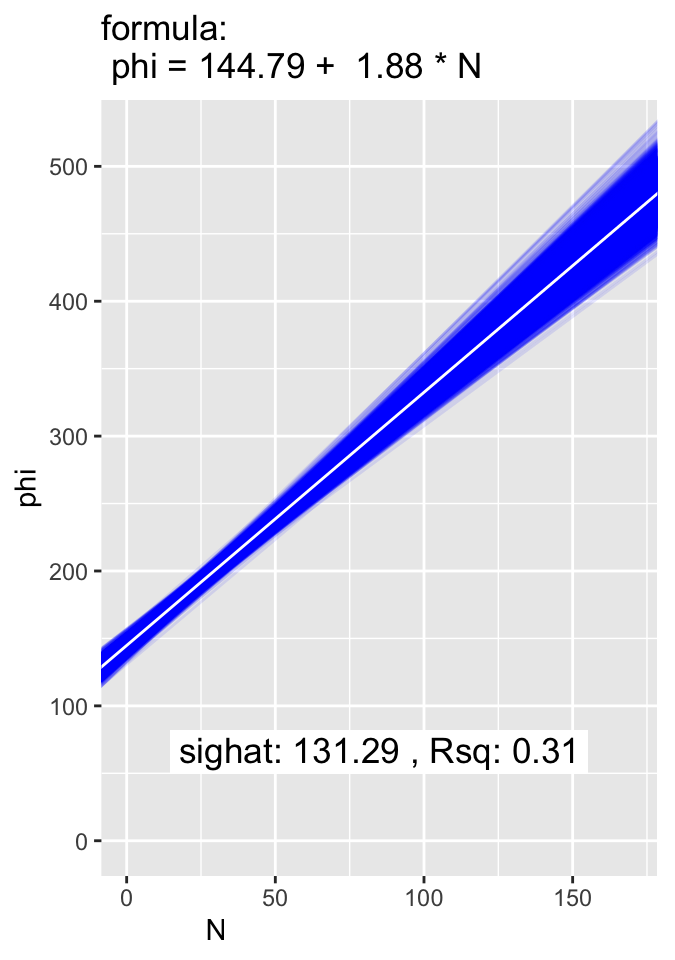}\\
      \includegraphics[width=2.125in]{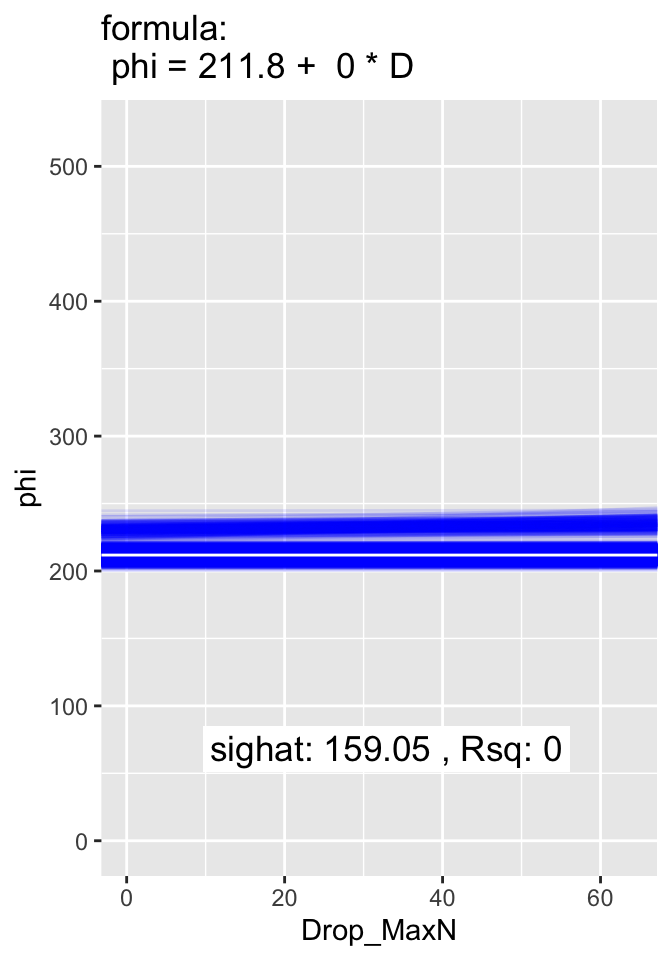} $\quad$
      \includegraphics[width=2.125in]{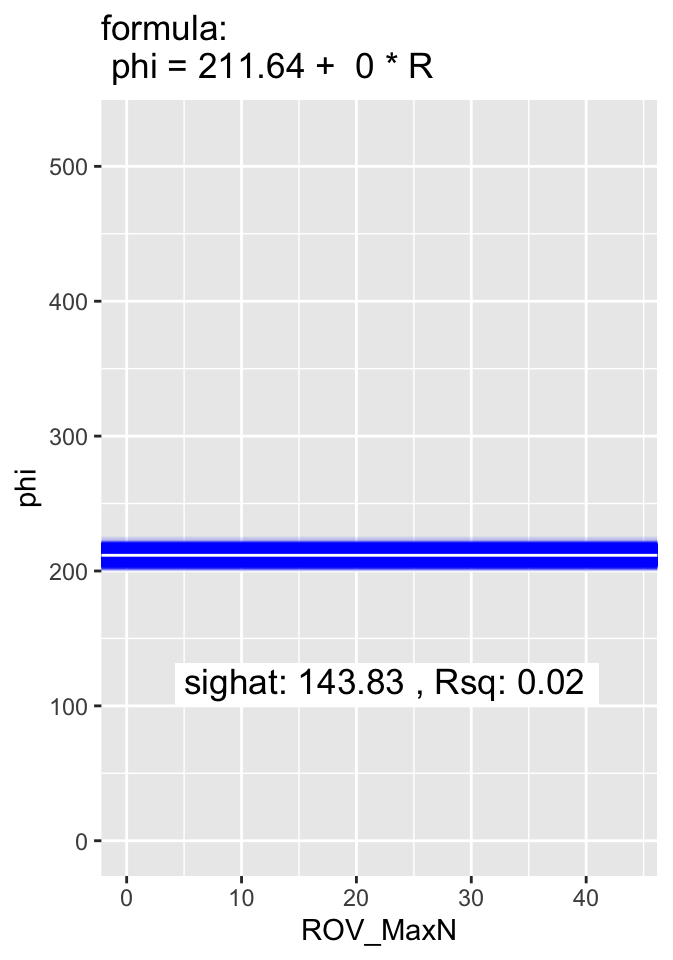}
}
\caption{Same as Figure~\ref{fig-MR3log} but for regression on the count scale. The horizontal regression lines in three of the panels are due to posterior draws of $\phi$ being negatively correlated with $N^{(mr)}, y_D,$ and $y_R$, thus producing constrained fits with a 0 slope.\label{fig-MR3}}
  \end{figure}
%
%
%
%
%
\newpage
\item \label{MR4} We incorporated both MR and the updated acoustic counts as paired observations of the true absolute abundance. The models we fitted here were based on the following:
      \begin{align*}
      \text{Level 1.1:} && y_{ijk\ell} &\sim \text{Pois}(\mu_{ijk\ell}) \\
      \text{Level 1.2:} && \text{for } h=1 \text{ (acoustic)}: \quad \quad N_{ijk} &\sim \text{Pois}(\tau_{hijk}) \quad \text{for all } i,j,k \\
      && \text{for } h=2 \text{ (mark-recapture)}: \quad\quad N_{i1k}^{(mr)} &\sim \text{Pois}(\tau_{hi1k}) \quad \text{for all } i,k \quad (j=1 \text{ only}) \\
      \text{Level 2.1:} && \log\mu_{ijk\ell} &= \nu_{y,i\ell} + \gamma_{y,j\ell} + \beta_{1,j\ell}\log\phi_{ijk} + \varepsilon_{ijk\ell} \\
      && \left[ \begin{matrix}
      \varepsilon_{ijkD} \\
      \varepsilon_{ijkS} \\
      \varepsilon_{ijkT} \\
      \varepsilon_{ijkR} 
      \end{matrix} \right] &\sim \text{MVN}(\mathbf{0}, \sigma_y^2\mathbb{P}) \\ 
      && \mathbb{P}_{\ell_1,\ell_2} &= \mathbb{P}_{\ell_2,\ell_1} = \rho \quad \text{ for all } \ell_1\ne\ell_2 \\
      && 0 = \sum_i \nu_{y,i\ell} &= \sum_j \gamma_{y,j\ell}\quad \text{for identifiability} \\
      \text{Level 2.2:} && \text{for } h=1: \quad \quad \log(r_{ijk}\tau_{hijk}) &= \log\phi_{ijk} + \xi_1 + \zeta_{hijk} \\
      && \text{for } h=2: \quad \quad \log\tau_{hi1k} &= \log\phi_{i1k} + \xi_2 + \zeta_{hi1k} \quad (j=1 \text{ only}) \\
      && \text{for } h=1,2: \quad \quad \zeta_{hi1k} &\sim \mathcal{N}(0, \sigma_{x,1}^2) \\
      && \text{for } h=1,2: \quad \quad \zeta_{hi2k} &= 0 \quad \text{for identifiability due to } k_{i2}\le 2 \\
      && 0 &= \sum_s \xi_s \quad \text{for identifiability} \\
      \text{Level 3:} && \log\phi_{ijk} &= \beta_0 + \nu_{x,i} + \gamma_{x,j} + \delta_{ijk} \\
      && \delta_{ijk} &\sim \mathcal{N}(0,\sigma_{\phi,j}^2) \\
      && 0 = \sum_i \nu_{x,i} &= \sum_j \gamma_{x,j} \quad \text{for identifiability}
      \end{align*}
    where \(r_{ijk}=1\) for all \(i,j,k\), i.e., we calibrated MR
    and acoustic counts against a latent ``true
    expectation'' (by letting each to be equal to the true GAJ
    density, \(\phi\), but subject to an offset on the log scale), and calibrated MaxN's against
    \(\phi\).
    
\textbf{RESULTS:} The parameter \(\log\phi\) barely
      correlated with logged MR counts, and had a non-unit
      slope with \(\log (N+1)\) (Figure~\ref{fig-MR4}).
      
    \begin{figure}[htbp]
\centering{
      \includegraphics[width=2.125in]{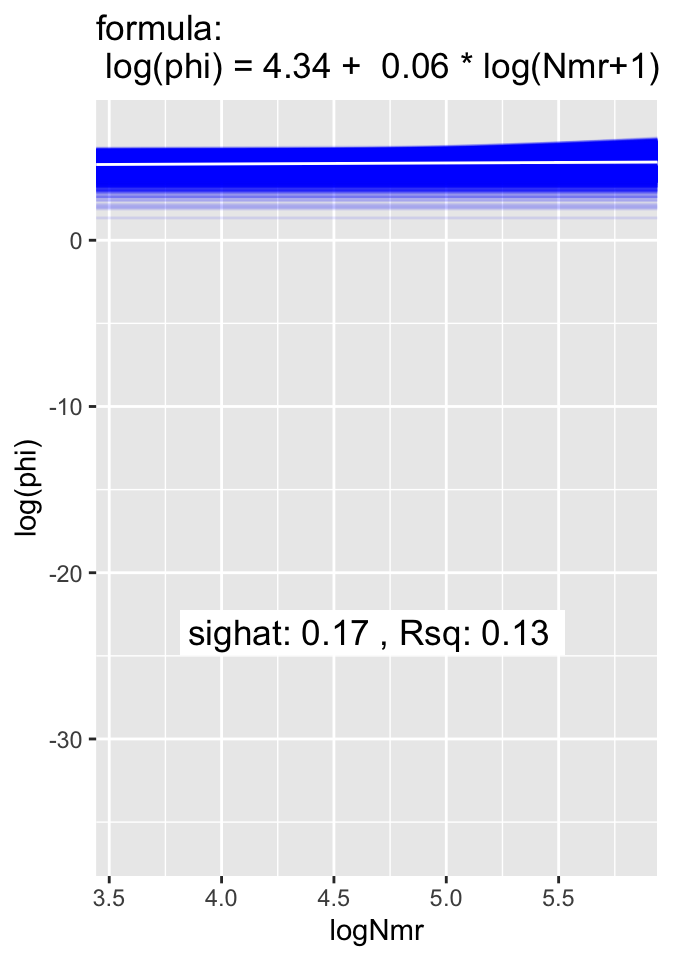}
      \(\quad\)
      \includegraphics[width=2.125in]{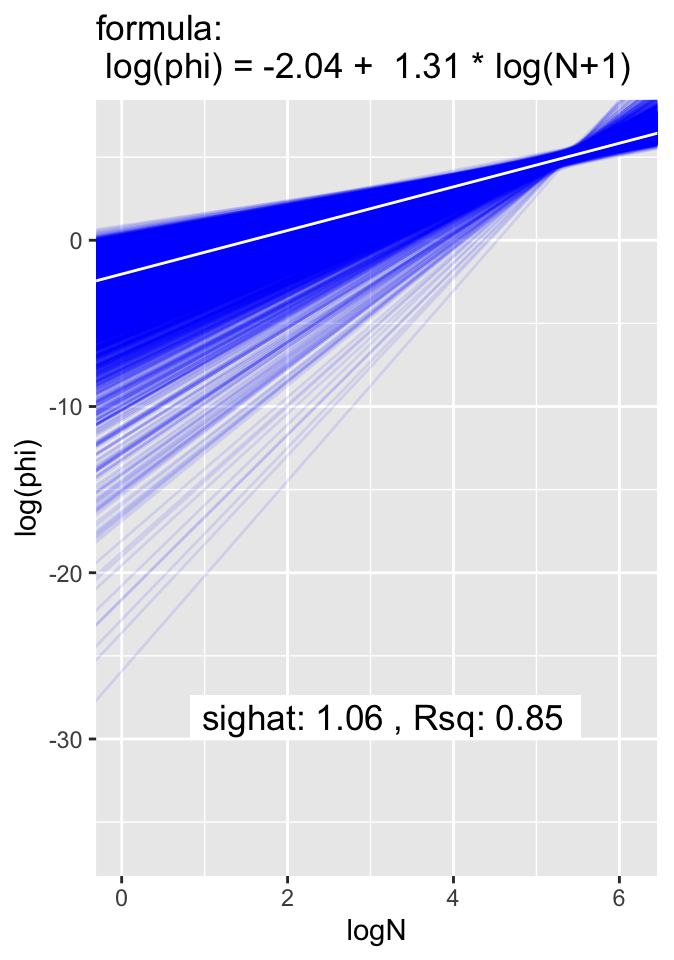}
}
\caption{Same as Figure~\ref{fig-MR2} but from models under Section~\ref{models-v2v3-mark-recpature-counts-gaj-only-and-raw-acoustic-counts-all-gaj-like-fish}, Item \#\ref{MR4}.\label{fig-MR4}}
    \end{figure}
\end{enumerate}

\subsection{Adjusted acoustic counts using preliminary scaling factor, $r^\ast$}\label{sec-modelV4}
These models employed a preliminary scaling factor, $r^\ast$, defined as the ratio between the MaxN for all \textit{Seriola} species (instead of only \textit{Seriola dumerili}, i.e., GAJ) and the MaxN for all GAJ+ species, pooled from drop- and ROV-camera observations only. At that stage of the modeling process, GAJ+ MaxN's were not yet available for the other two camera types. Also, we considered either \(N\) or \(N^{(f)}\) in the model based on the updated acoustic counts from Section~\ref{models-v2v3-mark-recpature-counts-gaj-only-and-raw-acoustic-counts-all-gaj-like-fish}, Item~\#\ref{MR3}.

The following model versions were fitted.

\vspace{5mm}

\begin{enumerate}
\item \label{rstar1} We calibrated GAJ MaxN's against the adjusted acoustic expected value, thus fitting model variants based on the following:
  \begin{align*}
  \text{Level 1.1:} && y_{ijk\ell} &\sim \text{Pois}(\mu_{ijk\ell}) \\
  \text{Level 2.1:} && N_{ijk} &\sim \text{Pois}(\phi_{ijk}/r^\ast_{ijk}) \quad \text{for all } i,j,k \\
  \text{Level 2.2:} && \log\mu_{ijk\ell} &= \nu_{y,i\ell} + \gamma_{y,j\ell} + \beta_{1,j\ell}\log\phi_{ijk} + \varepsilon_{ijk\ell} \\
  && \left[ \begin{matrix}
  \varepsilon_{ijkD} \\
  \varepsilon_{ijkS} \\
  \varepsilon_{ijkT} \\
  \varepsilon_{ijkR} 
  \end{matrix} \right] &\sim \text{MVN}(\mathbf{0}, \sigma_y^2\mathbb{P}) \\ 
  && \mathbb{P}_{\ell_1,\ell_2} &= \mathbb{P}_{\ell_2,\ell_1} = \rho \quad \text{ for all } \ell_1\ne\ell_2 \\
  && 0 = \sum_i \nu_{y,i\ell} &= \sum_j \gamma_{y,j\ell}\quad \text{for identifiability} \\
  \text{Level 3:} && \log\phi_{ijk} &= \beta_0 + \nu_{x,i} + \gamma_{x,j} + \delta_{ijk} \\
  && \delta_{ijk} &\sim \mathcal{N}(0,\sigma_{\phi,j}^2) \\
  && 0 = \sum_i \nu_{x,i} &= \sum_j \gamma_{x,j} \quad \text{for identifiability}
  \end{align*}

\textbf{RESULTS:}
Again, \(\log\phi\) barely correlated with
    logged MR counts, and had non-zero intercept and
    non-unit slope with \(\log [(N+1)r^\ast]\) (Figure~\ref{fig-rstar1}). However, results were desirable on the count scale (Figure~\ref{fig-rstar2}) for \(\phi\) vs.~each of \(N^{(mr)}\) (high correlation) and \(Nr^\ast\) (approximately 0 intercept and unit slope).  
\begin{figure}[htpb]
\centering{
    \includegraphics[width=2.125in]{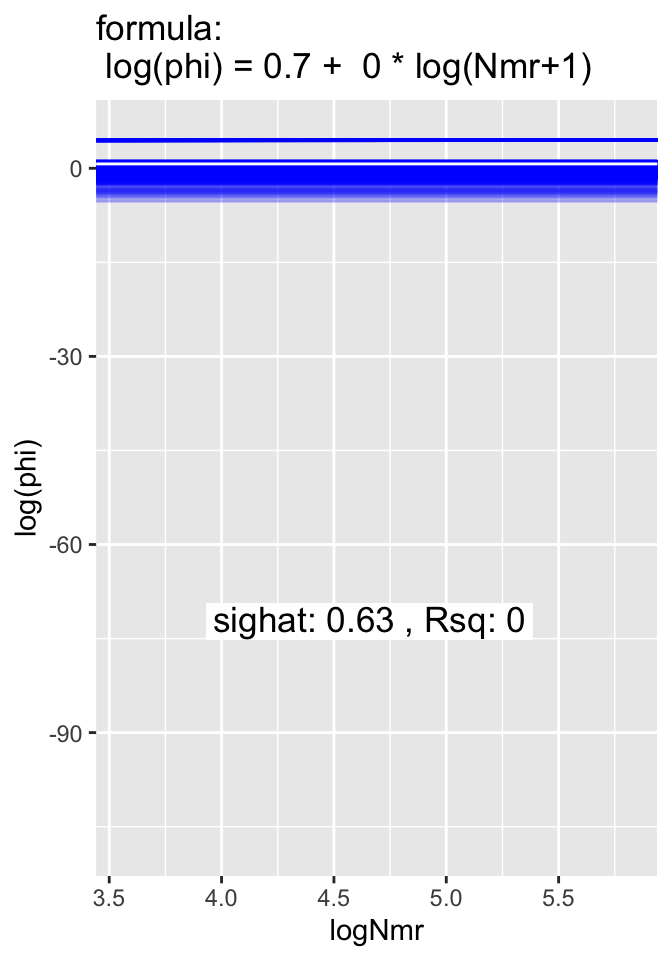}
    \(\quad\)
    \includegraphics[width=2.125in]{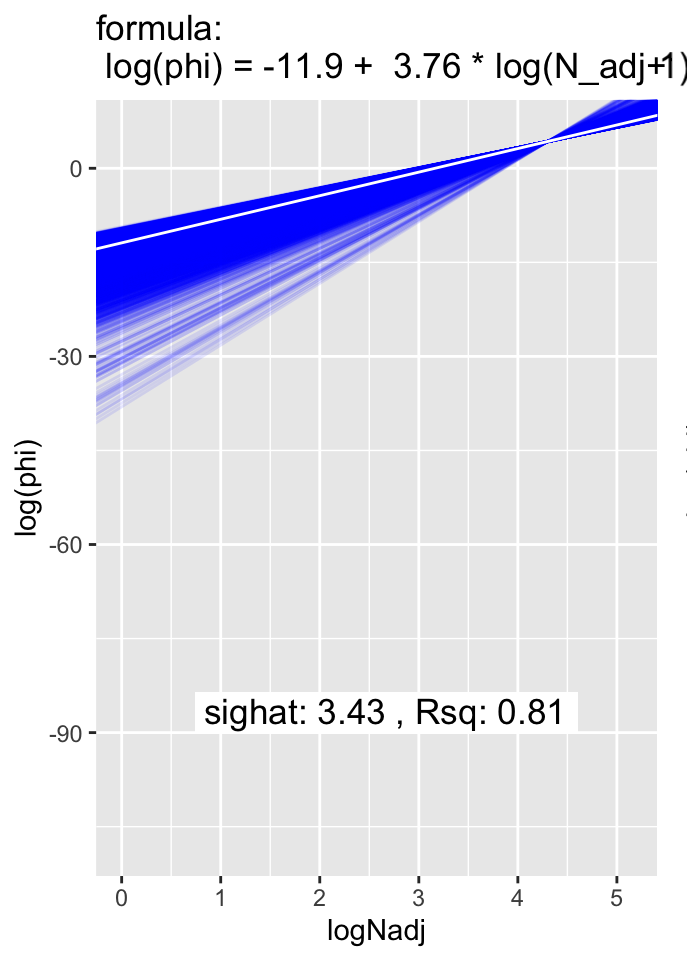}
}
\caption{Same as Figure~\ref{fig-MR2} but from models under Section~\ref{sec-modelV4}, Item \#\ref{rstar1}. `\texttt{N\_adj}' denotes $Nr^\ast$.\label{fig-rstar1}}
\end{figure}
\begin{figure}[htpb]
\centering{
    \includegraphics[width=2.125in]{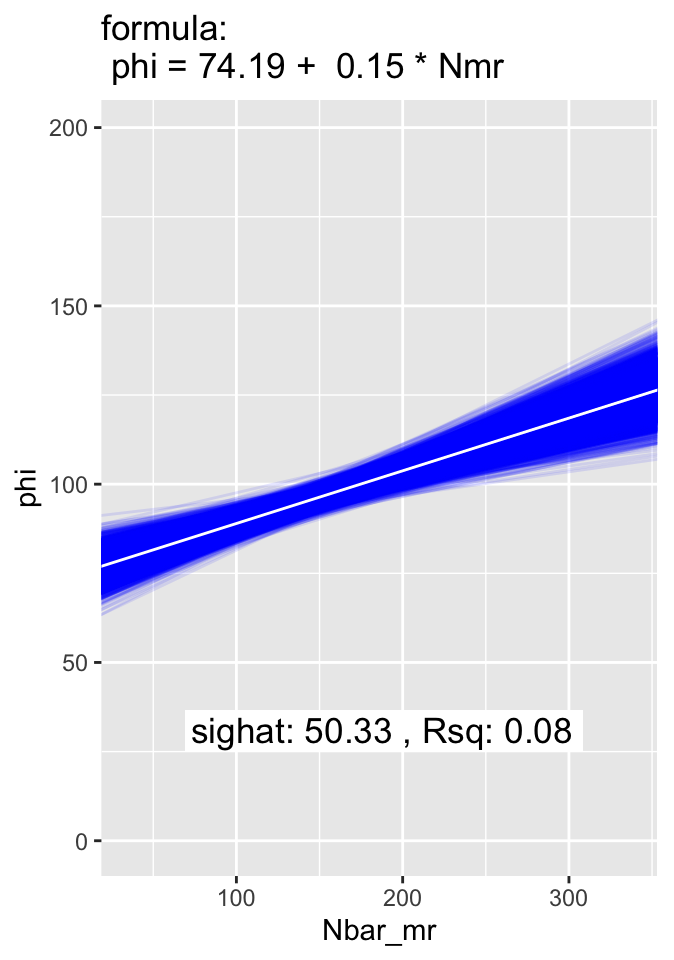} \(\quad\)
    \includegraphics[width=2.125in,]{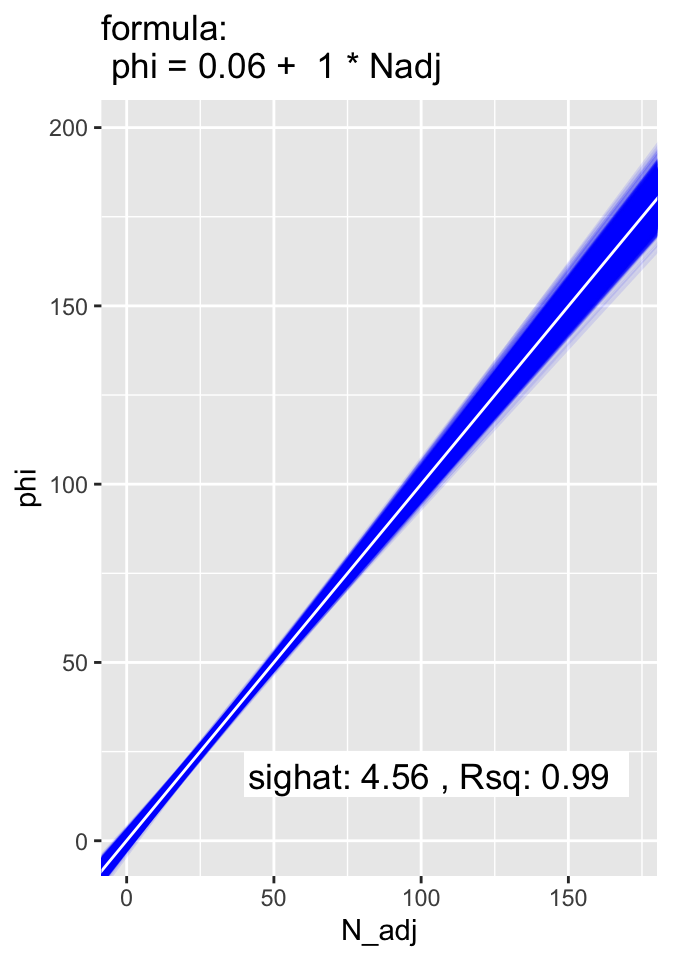}
}
\caption{Same as Figure~\ref{fig-rstar1} but for regression on the count scale.\label{fig-rstar2}}
\end{figure}
\vspace{5mm}

\item \label{rstar2} We replaced \(N\) in \#\ref{rstar1} in this section with \(N^{(f)}\) by replacing Level 2.1 with
\[
N_{ijk}^{(f)} \sim \text{Pois}(\phi_{ijk}/r^\ast_{ijk}) \quad \text{for all } i,j,k.
\]

\textbf{RESULTS:} Performance was less desirable than \#\ref{rstar1} (plots not shown).

\newpage
\item \label{rstar3} Similar to Item~\ref{MR4} under Section~\ref{models-v2v3-mark-recpature-counts-gaj-only-and-raw-acoustic-counts-all-gaj-like-fish}, here, we incorporated both \(Nr^\ast\) and \(N^{(mr)}\) as paired data on the true the absolute abundance, i.e., we fitted model variants based on the following:
  \begin{align*}
  \text{Level 1.1:} && y_{ijk\ell} &\sim \text{Pois}(\mu_{ijk\ell}) \\
  \text{Level 1.2:} && \text{for } h=1 \text{ (acoustic)}: \quad \quad N_{ijk} &\sim \text{Pois}(\tau_{hijk}) \quad \text{for all } i,j,k \\
  && \text{for } h=2 \text{ (mark-recapture)}: \quad\quad N_{i1k}^{(mr)} &\sim \text{Pois}(\tau_{hi1k}) \quad \text{for all } i,k \quad (j=1 \text{ only}) \\
  \text{Level 2.1:} && \log\mu_{ijk\ell} &= \nu_{y,i\ell} + \gamma_{y,j\ell} + \beta_{1,j\ell}\log\phi_{ijk} + \varepsilon_{ijk\ell} \\
  && \left[ \begin{matrix}
  \varepsilon_{ijkD} \\
  \varepsilon_{ijkS} \\
  \varepsilon_{ijkT} \\
  \varepsilon_{ijkR} 
  \end{matrix} \right] &\sim \text{MVN}(\mathbf{0}, \sigma_y^2\mathbb{P}) \\ 
  && \mathbb{P}_{\ell_1,\ell_2} &= \mathbb{P}_{\ell_2,\ell_1} = \rho \quad \text{ for all } \ell_1\ne\ell_2 \\
  && 0 = \sum_i \nu_{y,i\ell} &= \sum_j \gamma_{y,j\ell}\quad \text{for identifiability} \\
  \text{Level 2.2:} && \text{for } h=1: \quad \quad \log(r_{ijk}^\ast\tau_{hijk}) &= \log\phi_{ijk} + \xi_1 + \zeta_{hijk} \\
  && \text{for } h=2: \quad \quad \log\tau_{hi1k} &= \log\phi_{i1k} + \xi_2 + \zeta_{hi1k} \quad (j=1 \text{ only}) \\
  && \text{for } h=1,2: \quad \quad \zeta_{hi1k} &\sim \mathcal{N}(0, \sigma_{x,1}^2) \\
  && \text{for } h=1,2: \quad \quad \zeta_{hi2k} &= 0 \quad \text{for identifiability due to } k_{i2}\le 2 \\
  && 0 &= \sum_s \xi_s \quad \text{for identifiability} \\
  \text{Level 3:} && \log\phi_{ijk} &= \beta_0 + \nu_{x,i} + \gamma_{x,j} + \delta_{ijk} \\
  && \delta_{ijk} &\sim \mathcal{N}(0,\sigma_{\phi,j}^2) \\
  && 0 = \sum_i \nu_{x,i} &= \sum_j \gamma_{x,j} \quad \text{for identifiability}
  \end{align*}

\textbf{RESULTS:} The parameter \(\log\phi\) barely
    correlated with logged MR counts, and had a non-unit slope
    with \(\log [(N+1)r^\ast]\) (Figure~\ref{fig-rstar3}). Performance on the count scale was not evaluated.
\begin{figure}[htbp]
\centering{
    \includegraphics[width=2.125in]{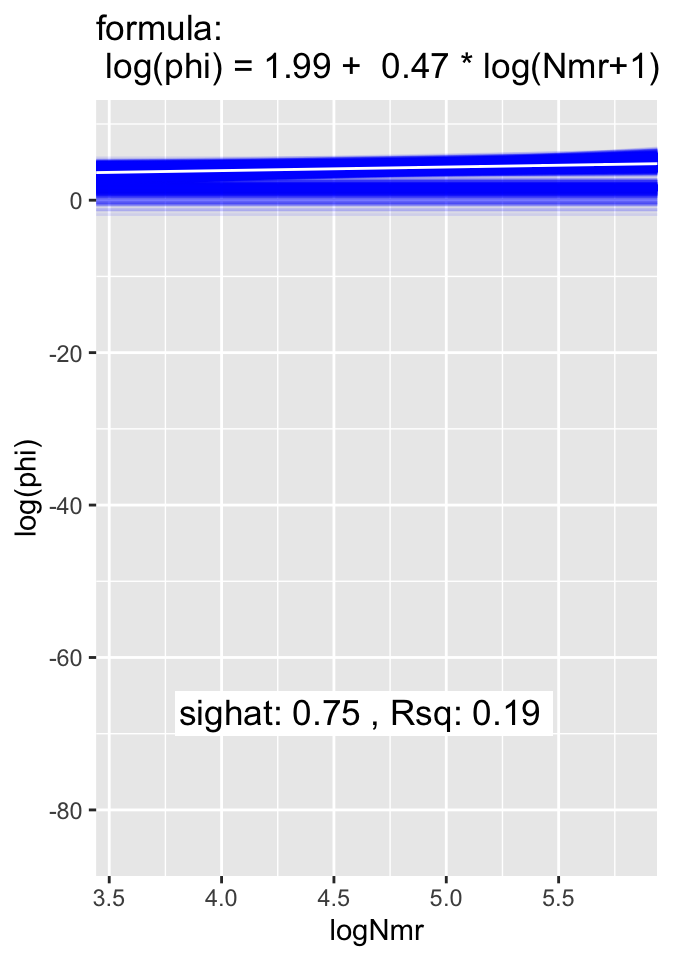}
    \(\quad\)
    \includegraphics[width=2.125in]{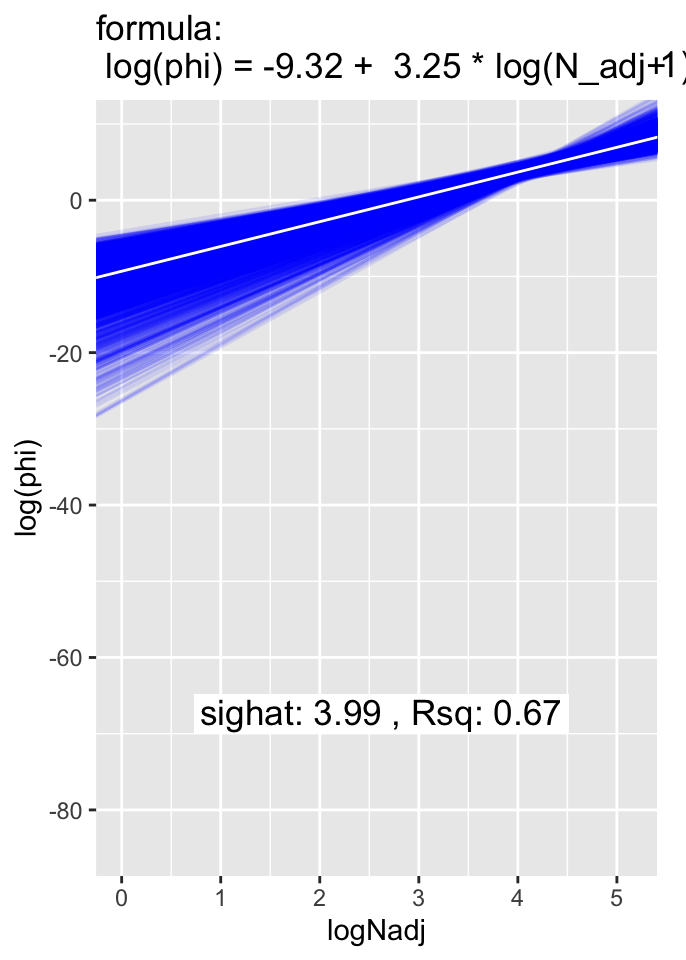}
}
\caption{Same as Figure~\ref{fig-rstar1} but from models under Section~\ref{sec-modelV4}, Item \#\ref{rstar3}.\label{fig-rstar3}}
\end{figure}
\vspace{5mm}

\item \label{rstar4} We repeated the models in \#\ref{rstar3} under this section, but replaced \(N\) with \(N^{(f)}\).  

\textbf{RESULTS:} Performance on the log scale was less desirable than \#\ref{rstar3} (plots not shown). Again, performance on the count scale was not evaluated.
\end{enumerate}

\subsection{Adjusted acoustic counts by pooling all camera ratios}\label{sec-modelV5}

At this point, it was decided that $N^{(mr})$ should no longer be modeled alongside $N$, but should be used purely for model validation. The decision was made due to the poor performance of models that included $N^{(mr)}$, but also because $N^{(mr)}$ was only available at one type of habitat (``super pyramid'') and thus may be inherently biased.

Seeing that \#\ref{rstar1} under Section~\ref{sec-modelV4} was promising, we proceeded to repeat \#\ref{rstar1}--\ref{rstar2} from that section, but replacing \(r^\ast\) with \(r\) (based on all four cameras).

We fitted model variants based on the following:
\begin{align*}
\text{Level 1.1:} && y_{ijk\ell} &\sim \text{Pois}(\mu_{ijk\ell}) \\
\text{Level 2.1:} && N_{ijk} &\sim \text{Pois}(\phi_{ijk}/r_{ijk}) \quad \text{for all } i,j,k \\
\text{Level 2.2:} && \log\mu_{ijk\ell} &= \nu_{y,i\ell} + \gamma_{y,j\ell} + \beta_{1,j\ell}\log\phi_{ijk} + \varepsilon_{ijk\ell} \\
&& \left[ \begin{matrix}
\varepsilon_{ijkD} \\
\varepsilon_{ijkS} \\
\varepsilon_{ijkT} \\
\varepsilon_{ijkR} 
\end{matrix} \right] &\sim \text{MVN}(\mathbf{0}, \sigma_y^2\mathbb{P}) \\ 
&& \mathbb{P}_{\ell_1,\ell_2} &= \mathbb{P}_{\ell_2,\ell_1} = \rho \quad \text{ for all } \ell_1\ne\ell_2 \\
&& 0 = \sum_i \nu_{y,i\ell} &= \sum_j \gamma_{y,j\ell}\quad \text{for identifiability} \\
\text{Level 3:} && \log\phi_{ijk} &= \beta_0 + \nu_{x,i} + \gamma_{x,j} + \delta_{ijk} \\
&& \delta_{ijk} &\sim \mathcal{N}(0,\sigma_{\phi,j}^2) \\
&& 0 = \sum_i \nu_{x,i} &= \sum_j \gamma_{x,j} \quad \text{for identifiability}
\end{align*}

\textbf{RESULTS:} Focusing on the count scale, we saw the best overall
  alignment of $\phi$ with \(N^{(mr)}, rN\), and \(rN^{(f)}\) among all fitted models throughout the Appendix~\ref{sec-appdx-timeline} timeline up until this point. 

\subsection{Adjusted acoustic counts by updating pooled camera ratios}\label{sec-modelV6}

At this point, it was decided that the adjustment factor $r$ would be recomputed to align with how they would be computed from the field survey data. Additionally, as the approach under Section \ref{sec-modelV5} was the most promising, we reintroduced $\beta_{y,0,\ell}$ to Level 2.2 before proceeding. The model variant with the best performance under this section is described in Section~\ref{sec-final-model}.

\section{GAJ+ species}\label{sec-GAJ+}

\begin{table}[htbp]
\caption{\textbf{``Large-bodied'' species list.} A `1' under a camera type denotes a species that was observed in the experiment by that camera type, and a `1' under the
``Ratio'' column denotes a large-bodied fish (total length above 20 cm)
that generally are found above 0.5 m from the seafloor, do not form
dense aggregations (i.e., are solitary or form loosely packed schools),
and are thus considered to be acoustically indistinguishable from GAJ.\label{tab-GAJ+}}
\begin{center}
\begin{tabular}{llrrrrr}
  \hline
Scientific Name & Common Name & Drop & SBRUV & Trap & ROV & Ratio\\
  \hline
\em{Lutjanus campechanus} & Red Snapper & 1 & 1 & 1 & 1 & 1\\
\em{Seriola dumerili} & Greater Amberjack & 1 & 1 & 1 & 1 & 1\\
\em{Seriola rivoliana} & Almaco Jack & 1 & 1 & 1 & 1 & 1\\
\em{Balistes capriscus} & Grey Triggerfish & 1 & 1 & 1 & 1 & 1\\
\em{Thunnus atlanticus} & Blackfin Tuna & 1 & 0 & 0 & 0 & 1\\
\em{Euthynnus alletteratus} & Little Tunny & 1 & 1 & 1 & 0 & 1\\
\em{Rhomboplites aurorubens} & Vermilion Snapper & 1 & 1 & 1 & 1 & 0\\
\em{Centropristis ocyurus} & Bank Sea Bass & 1 & 1 & 1 & 1 & 0\\
\em{Carcharodon carcharias} & Great White Shark & 1 & 1 & 1 & 1 & 1\\
\em{Chaetodipterus faber} & Atlantic Spadefish & 1 & 1 & 1 & 1 & 1\\
\em{Lutjanus griseus} & Grey Snapper & 1 & 1 & 1 & 1 & 1\\
\em{Pterois volitans} & Lionfish & 0 & 1 & 1 & 1 & 0\\
\em{Lutjanus synagris} & Lane Snapper & 0 & 1 & 1 & 1 & 1\\
\em{Holacanthus bermudensis} & Blue Angelfish & 0 & 1 & 1 & 1 & 0\\
\em{Chaetodon spp.} & Butterflyfish & 0 & 0 & 1 & 1 & 0\\
\em{Chaetodon ocellatus} & Spotfin Butterflyfish & 0 & 1 & 0 & 1 & 0\\
\em{Rypticus maculatus} & Whitespotted Soapfish & 0 & 1 & 0 & 1 & 0\\
\em{Pomacentridae} & Damselfishes & 0 & 0 & 0 & 1 & 0\\
\em{Labridae} & Wrasses & 0 & 0 & 0 & 1 & 0\\
\em{Mycteroperca phenax} & Scamp & 0 & 0 & 1 & 1 & 1\\
\em{Aluterus monoceros} & Unicorn Filefish & 0 & 1 & 1 & 1 & 1\\
\em{Pareques spp.} & Pareques spp. & 0 & 0 & 0 & 1 & 0\\
\em{Archosargus probatocephalus} & Sheepshead & 0 & 1 & 0 & 1 & 1\\
\em{Haemulon aurolineatum} & Tomtate & 0 & 1 & 1 & 1 & 0\\
\em{Stenotomus caprinus} & Longspine Porgy & 0 & 1 & 1 & 1 & 0\\
\em{Calamus spp.} & Porgies & 0 & 0 & 1 & 1 & 1\\
\em{Rachycentron canadum} & Cobia & 0 & 0 & 1 & 1 & 1\\
\em{Eques lanceolatus} & Jack-knife Fish & 0 & 0 & 0 & 1 & 0\\
\em{Neopomacentrus cyanomos} & Regal Demoiselle & 0 & 0 & 0 & 1 & 0\\
\em{Heteroconger sp.} & Garden Eel & 0 & 1 & 0 & 0 & 0\\
\em{Carcharhinus plumbeus} & Sandbar Shark & 0 & 1 & 1 & 0 & 1\\
\em{Echeneis spp.} & Sharksucker & 0 & 1 & 1 & 0 & 1\\
\em{Diplectrum formosum} & Sand Perch & 0 & 1 & 0 & 0 & 0\\
\em{Centropristis philadelphica} & Rock Sea Bass & 0 & 1 & 0 & 0 & 0\\
\em{Orthopristis chrysoptera} & Pigfish & 0 & 1 & 0 & 0 & 0\\
\em{Lagodon rhomboides} & Pinfish & 0 & 1 & 0 & 0 & 0\\
\em{Calamus leucosteus} & Whitebone Porgy & 0 & 1 & 1 & 0 & 1\\
\em{Scomberomorus spp.} & Spanish Mackerel & 0 & 1 & 1 & 0 & 1\\
\em{Caranx crysos} & Blue Runner & 0 & 0 & 1 & 0 & 1\\
\em{Carcharhinus leucas} & Bull Shark & 0 & 0 & 1 & 0 & 1\\
\em{Pagrus pagrus} & Red Porgy & 0 & 0 & 1 & 0 & 1\\
\em{Mycteroperca microlepis} & Gag & 0 & 0 & 1 & 0 & 1\\
\em{Anisotremus virginicus} & Porkfish & 0 & 0 & 1 & 0 & 1\\
\em{Sphyraena barracuda} & Great Barracuda & 0 & 0 & 1 & 0 & 1\\
\em{Selene vomer} & Lookdown & 0 & 0 & 1 & 0 & 1\\
\em{Caranx hippos} & Crevalle Jack & 0 & 0 & 1 & 0 & 1\\
\em{Aluterus scriptus} & Scrawled Filefish & 0 & 0 & 1 & 0 & 1\\
\em{Calamus nodosus} & Knobbed Porgy & 0 & 0 & 1 & 0 & 1\\
\em{Pristipomoides sp.} & Wenchmen & 0 & 0 & 1 & 0 & 0\\
\em{Bodianus pulchellus} & Spotfin Hogfish & 0 & 0 & 1 & 0 & 1\\
\em{Epinephelus morio} & Red Grouper & 0 & 0 & 1 & 0 & 1\\
\em{Decapterus punctatus} & Round Scad & 0 & 0 & 1 & 0 & 0\\
   \hline
\end{tabular}
\end{center}
\end{table}

As shown in Figure~\ref{fig-camratios}, there is a great deal of noise
in camera-specific ratios relative to the small sample sizes. Much of
this noise is due to the highly varying detectability across
camera-species combinations as can be seen in Table~\ref{tab-GAJ+}. For example,
the narrow field of view and stationary qualities of the drop camera
make individuals less detectable compared to the ROV, particularly for
those species that tend to be closely associated with the reef
structure. This variability translates into differences in the
denominator of the camera-specific scaling factor.

\section{Model-based derivation of fish density using acoustic counts}\label{sec-appdx-acoustic}

\begin{figure}[htpb]
\includegraphics[width=\textwidth]{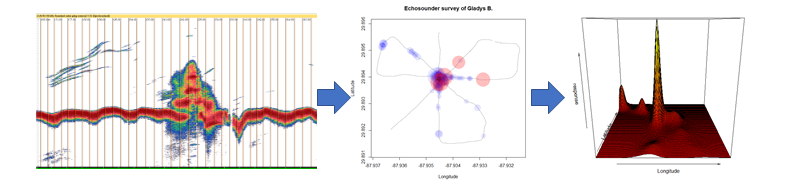}
\caption{\label{fig-interp}An example of acoustic signals recorded within a grid cell. Left panel: raw signals. Middle panel: spatial zone identified by the acoustic signal processing algorithm as the domain relevant to fish activities. Right panel: spatial interpolation of fish counts based on the middle panel.}
\end{figure}

Acoustic echosounder data provide geo-referenced counts for GAJ+ species along
  multiple transects around a reef structure (Figure~\ref{fig-interp}, left panel). Ideally, model-based spatial interpolation is applied to infer fish counts across the entire sampled area made up of many grid cells, each being an observational unit for the study-wide GAJ survey. Model-based interpolation also allows uncertainty quantification associated with such inference.
  
If a habitat structure is part of a large contiguous reef structure or unconsolidated bottom with non-trivial spatial footprint, then the average height of the interpolated surface over the total footprint of all such areal structures within the $g$th grid cell can be an estimate of fish density at such structures in grid cell $g$. In contrast, if the habitat structure is a point in space with minimal footprint (e.g., most artificial reefs), one should identify a \textit{region of association} around the point-level structure to create an approximate footprint, where the association refers to the footprint of fish activity due to its attraction to the point-level reef structure in question. Identification may be possible from the shape of the fitted surface over cell $g$ (Figure~\ref{fig-interp}, right panel). An ad-hoc alternative method is to pre-assign buffer zones (e.g.,~100m radius) around each point-level structure to create an approximate region of association.

\end{appendix}


\begin{acks}[Acknowledgments]
GSC acknowledges William \& Mary Research Computing for providing computational resources and technical support that have contributed to the results reported within this paper. URL: \url{https://www.wm.edu/it/rc}.  
\end{acks}

\begin{funding}
This work is part of the Greater Amberjack Research funded by the Mississippi-Alabama Sea Grant Consortium through the NOAA Award NA20OAR4170494. URL: \url{https://masgc.org/greater-amberjack}
\end{funding}


\bibliographystyle{imsart-nameyear} 
\bibliography{chiu-et-al2025}       

\end{changemargin}
\end{document}